\documentclass[10pt,amsmath,aps,floatfix,nofootinbib,notitlepage,prd,superscriptaddress,twocolumn]{revtex4-1}

\usepackage[T1]{fontenc}
\usepackage[usenames,dvipsnames]{xcolor}
\usepackage{aas_macros,graphicx,hyperref,orcidlink}
\hypersetup{colorlinks=true,citecolor=RoyalBlue,linkcolor=RoyalBlue,urlcolor=RoyalBlue}

\let\originalleft\left
\let\originalright\right
\renewcommand{\left}{\mathopen{}\mathclose\bgroup\originalleft}
\renewcommand{\right}{\aftergroup\egroup\originalright}

\newcommand{\ab}[1]{\left|#1\right|}
\newcommand{\av}[1]{\left\langle#1\right\rangle}
\newcommand{\br}[1]{\left[#1\right]}
\newcommand{\cu}[1]{\left\{#1\right\}}
\newcommand{\pa}[1]{\left(#1\right)}
\newcommand{\ed}{\mathop{}\!\mathrm{d}}
\DeclareMathOperator\arcsinh{arcsinh}

\begin{document}

\title{Assessing the impact of instrument noise and astrophysical fluctuations on measurements of the first black hole photon ring}

\author{Alejandro C\'ardenas-Avenda\~no\,\orcidlink{0000-0001-9528-1826}} 
\email{cardenas-avendano@princeton.edu}
\affiliation{Princeton Gravity Initiative, Princeton University, Princeton, New Jersey 08544, USA}
\affiliation{Department of Physics, Princeton University, Princeton, New Jersey 08544, USA}

\author{Lennox Keeble\,\orcidlink{0009-0009-5796-631X}} 
\affiliation{Department of Physics, Princeton University, Princeton, New Jersey 08544, USA}

\author{Alexandru Lupsasca\,\orcidlink{0000-0002-1559-6965}}
\affiliation{Department of Physics \& Astronomy, Vanderbilt University, Nashville, Tennessee 37212, USA}

\begin{abstract}
Currently envisioned extensions of the Event Horizon Telescope to space will soon target the black hole photon ring: a narrow ring-shaped imprint of a black hole's strong gravity produced in its images by highly bent photon trajectories.
In principle, the shape of the photon ring encodes information about the geometry of the underlying black hole spacetime.
In practice, however, whether or not this information can be extracted from the ring shape depends on several factors, ranging from the astrophysical details of the emitting source (such as the magnitude of its plasma fluctuations) to the specific configuration of the interferometric array (such as the separation between its telescopes, or the level of noise in its instruments).
Here, we employ a phenomenological model to assess the impact of astrophysical fluctuations and instrument noise on the inferred shape of the photon ring.
Our systematic study of several astrophysical profiles suggests that this shape can be measured even in the presence of instrument noise across a wide range of baselines.
The measurement accuracy and precision appear relatively insensitive to the noise level, up to a sharp threshold beyond which any measurement becomes incredibly challenging (at least without recourse to more sophisticated data analysis methods).
Encouragingly, we find that only a few snapshot images are generally needed to overcome the impact of astrophysical fluctuations and correctly infer the ring diameter.
Inference becomes more challenging when analyzing the visibility amplitude in a baseline window that is not entirely dominated by a single photon ring.
Nevertheless, in most cases, it is still possible to fit a ring shape with the correct fractional asymmetry.
These results provide excellent prospects for future precision measurements of black hole spin and fundamental astrophysics via black hole imaging.
\end{abstract}

\maketitle

\section{Introduction}

Although the individual pulses of a pulsar can display significant variability, their average is remarkably stable.
Pulsar timing array experiments rely on this stability to test a wide range of relativistic predictions~\cite{Verbiest2021,NANOGrav2023}.
While images of a supermassive black hole can likewise exhibit large variability, their time average is also expected to be highly stable: individual snapshots may display plasma flares or other prominent transient features, but such source fluctuations ought to wash out in the long run, leaving an image dominated by the signature of the black hole---its photon ring~\cite{JohnsonLupsasca2020,GLM2020}.
This ring is a narrow image feature produced by photons on highly bent trajectories that orbited the black hole multiple times on their way from source to observer.
It consists of multiple subrings, each of which shows a lensed image of the main emission, indexed by the number $n$ of photon half-orbits executed around the black hole~\cite{Gralla2019,GrallaLupsasca2020a,Lupsasca2024}.
These subrings encircle the central brightness deficit caused by the event horizon and form the part of the image that belongs to the black hole itself, rather than its surrounding emission: the photon ring is a persistent ``stamp'' on the image of the black hole that carries information about its strong gravity, encoded in the shape of the subrings.
Planning is now underway for a space mission capable of performing interferometric observations on baselines long enough to resolve the first ($n=1$) rings of the black holes M87* and Sgr~A*~\cite{Gurvits2022,Kurczynski2022,Kudriashov2021}.

As this experimental effort unfolds, ongoing theoretical work~\cite{JohnsonLupsasca2020,Gralla2020,GrallaLupsasca2020c,GLM2020,Paugnat2022,Vincent2022,CardenasAvendano2023} has focused on deriving predictions for the interferometric signature of the photon ring.
By now, multiple studies have investigated how feasible it might be to extract this signature from time-averaged black hole images.
Initially, the focus was on the $n=2$ photon ring, which produces a very clean signal~\cite{GLM2020,Paugnat2022,Vincent2022}, but it has recently shifted to the $n=1$ ring \cite{CardenasAvendano2023}, as it is more easily accessible observationally and will hence be resolved first.

Deriving a sharp, general-relativistic prediction for the interferometric signature of the $n=1$ ring is nontrivial.
The reason is that, strictly speaking, general relativity only completely fixes the shape of the limiting $n\to\infty$ ring: an infinitely thin ring known as the ``critical curve'' $\tilde{\mathcal{C}}$~\cite{Bardeen1973}.
Although its angle-dependent diameter $\tilde{d}_\varphi$ follows an analytically known form~\cite{GrallaLupsasca2020c}, this mathematical shape is not in itself observable.
What is in principle observable are the finite-$n$ rings, which converge exponentially fast in $n$ to $\tilde{\mathcal{C}}$, but can nonetheless display corrections in $1/n$ that remain significant at $n=1$.

In particular, the first $n=1$ photon ring has a larger width-to-diameter ratio than subsequent rings---of order $\sim\!10\%$---and therefore lacks a sharply defined diameter in the image domain.
Nevertheless, it \emph{does} admit a sharp \textit{interferometric} diameter $d_\varphi^{(1)}$, which can be defined in the frequency domain by the periodic ringing that the ring produces in the radio visibility \cite{CardenasAvendano2023}.
In summary, even though general relativity (GR) only directly determines a functional form for the angle-dependent diameter $\tilde{d}_\varphi$ of $\tilde{\mathcal{C}}$, this very same functional form also constrains the interferometric ring diameter $d_\varphi^{(1)}$ of the first photon ring.

Although these theoretical findings are encouraging, in practice, photon ring measurements face several hurdles.
The main challenge arises from the very narrow width of the rings: at our present observing frequencies, resolving a photon ring requires very-long-baseline interferometric (VLBI) observations on baselines extending from ground stations all the way to a radio antenna in space~\cite{JohnsonLupsasca2020,Tiede2022,Lockhart2022}.

Besides the considerable technological advances needed to carry out such observations in space, the narrow width of the photon ring also poses challenges to our existing modeling approaches: to adequately resolve it, one has to ray trace general-relativistic magnetohydrodynamic (GRMHD) simulations of the accretion flow onto the black hole at very high image resolutions (for more details on these types of simulations, see, e.g., Refs.~\cite{Font2008,Porth2019,Wong2022,Mizuno2022}). 

Such time-averaged GRMHD simulations have shown that photon rings are persistent, sharp features that come to dominate VLBI observations after averaging them over sufficiently long timescales~\cite{JohnsonLupsasca2020}.
Because of their onerous computational cost, however, these simulations cannot explore the vast space of plausible configurations, which is parameterized by many variables, including: black hole spin, observer inclination, the plasma heating model, or accretion flow magnetization states, among many others.

Properly accounting for the time variability of sources presents another challenge: for instance, the variability observed in Sgr~A*, the supermassive black hole at the center of our galaxy, is a factor of two smaller than the source-integrated variability predicted by our GRMHD models~\cite{EHT2022e,Wielgus2022}.
In other words, we expect photon rings to appear in time-averaged images, but the exact details of how this averaging is to be carried out remain unclear.

For these reasons, prior studies of the photon ring and its interferometric signature have mostly revolved around time-averaged images~\cite{Gralla2019,JohnsonLupsasca2020,GLM2020,Paugnat2022,Vincent2022,CardenasAvendano2023}.
To forecast future space-VLBI experiments, however, one must tackle the time variability head-on, and do so at high resolution.
In addition, accurate forecasting must take into account the fine details of the interferometric array: in particular, one must choose an orbit for the space telescope, and model the physical and operational constraints imposed by the spacecraft design and mission architecture~\cite{Hudson2023}.
If, for example, the space dish observes with the frequencies of existing ground stations (namely, $86$, $230$, or $345$\,GHz), then its orbit determines the baseline coverage (i.e., the subset of the frequency domain) accessible to the array.

Here, we begin to incorporate these complications into our forecasting, using a phenomenological model built into the Adaptive Analytical Ray Tracing code \texttt{AART}~\cite{CardenasAvendano2022}.
This code exploits the integrability of light propagation in the Kerr spacetime to produce high-resolution black hole images and their corresponding visibilities, as would be observed on long space-to-ground baselines.
We conduct a survey of parameter space to determine how accurately the shape of the first photon ring can be inferred in the presence of various levels of instrument noise and source variability.
The survey also varies baseline lengths and ranges as a proxy for the yet-to-be-pinned-down orbit.

To model instrument noise, we add complex Gaussian noise to the raw, simulated radio visibility.
Meanwhile, we model astrophysical fluctuations using \texttt{inoisy}~\cite{Lee2021}, a code capable of replicating many features of accretion processes by generating realizations of Gaussian random fields (GRFs) with tuneable spatiotemporal correlations.
That is, rather than strictly adhering to a physics-based simulation, such as in GRMHD, we instead resort to a stochastical model of fluctuations that allows us to vary their power and correlation structure systematically.
This enables us to parameterize our uncertainty about the scale of the fluctuations, and to thereby offer a broad understanding of the various factors that will influence future observations.
Since the amount of time-averaging required for a future space-VLBI mission to isolate the photon ring ultimately depends on the variability of the underlying plasma, we consider multiple astrophysical models with different variability and emission geometries.

The results of our analysis suggest that, even in the presence of substantial levels of \emph{both} instrument noise and astrophysical fluctuations, it is possible to detect in the visibility amplitude a periodicity that encodes the shape of the first photon ring.
We are able to infer this shape with an accuracy and precision that is not significantly affected by the level of instrument noise, up to a certain, rather sharp threshold.
Detecting any signal becomes very challenging past this threshold, which therefore sets a clear limit on our measurement ability within this class of models and given our inference technique.
Averaging over multiple snapshots can increase this threshold and restore our ability to detect a photon ring signature even in the presence of elevated noise or stronger fluctuations. 

When analyzing the visibility amplitude in a baseline window where neither the first nor the second photon ring dominate the signal, it becomes more challenging to infer a periodicity and hence an interferometric diameter.
In such transition regimes, the periodic ringing of the visibility sometimes leads one to infer a ring shape that only differs from the ``correct’' shape (the one inferred in the absence of noise and fluctuations) by an overall scaling factor.
Fortuitously, this means that one can still infer the correct fractional asymmetry of the ring, even when its inferred diameter is off by some (small) factor.

The remainder of this paper is organized as follows.
In Sec.~\ref{sec:Model}, we briefly present the phenomenological model we use to simulate time-averaged and time-dependent data, and the inference scheme we use to extract ring diameters from a visibility.
Then, in Sec.~\ref{sec:InstrumentNoise}, we study the impact of instrument noise on time-averaged images.
In Sec.~\ref{sec:AstrophysicalFluctuations}, we move on to time-dependent models and study the effects of astrophysical fluctuations in isolation (in the absence of instrument noise).
Next, in Sec.~\ref{sec:NoiseAndFluctuations}, we study a highly challenging case in the presence of both instrument noise and astrophysical fluctuations, and demonstrate that a shape measurement still works under adverse conditions. Finally, we conclude with a summary of our results in Sec.~\ref{sec:Conclusion}, relegating some details of threshold selection and model characterization to two appendices.

\section{Phenomenological source model}
\label{sec:Model}

Throughout this paper, we use natural units with $G_{\rm N}$ and $c$ set to unity.
We simulate black hole images using the method of Ref.~\cite{GLM2020}, as implemented in \texttt{AART}~\cite{CardenasAvendano2022}.
In this section, we briefly review the key ingredients of this model, referring the reader to Ref.~\cite{CardenasAvendano2022} for further details. 

We consider a disk of emitters on circular-equatorial Keplerian orbits.
The disk terminates at an inner edge that corresponds to the innermost stable circular orbit (ISCO) radius.
Past it, the emitters plunge into the hole following Cunningham's prescription~\cite{Cunningham1975}.
The intensity at Cartesian image-plane position $(\alpha,\beta)$ is computed by (analytically) tracing the corresponding light ray back into the emitting region.
The observed intensity, $I_{\rm o}(\alpha,\beta)$, increases each time the ray intersects the accretion disk, by an amount determined by the source emissivity, $I_{\rm s}$.
The total observed intensity is then given by~\cite{CardenasAvendano2022}
\begin{align}
	\label{eq:ObservedIntensity}
	I_{\rm o}(\alpha,\beta)=\sum_{n=0}^{N-1}\zeta_ng^3\pa{r_{\rm s}^{(n)},\alpha,\beta}I_{\rm s}\pa{r_{\rm s}^{(n)},\phi_{\rm s}^{(n)}, t_{\rm s}^{(n)}},
\end{align}
where $x_{\rm s}^{(n)}=x_{\rm s}^{(n)}(\alpha,\beta)$ denotes the (analytically known) equatorial position where the ray intersects the equatorial plane for the $(n+1)^\text{th}$ time on its backward trajectory from image-plane position $(\alpha,\beta)$, up to a total number $N(\alpha,\beta)$ along its maximal extension.
The redshift factor $g$ is determined by the motion of the emitters, and $\zeta_n$ is a ``fudge'' factor, assumed to be equal to $1$ for $n=0$ and $3/2$ for $n\geq1$.
This factor is included to account for the effects of the disk's geometrical thickness~\cite{Chael2021,Vincent2022}. 

Under this prescription, the observed intensity receives two contributions: one purely geometric (encoded in the redshift factor $g$), and the other purely astrophysical (set by the source emissivity $I_{\rm s}$ and fudge factor $\zeta_n$).
It is now easy to prescribe time-averaged equatorial models, as in Refs.~\cite{GLM2020,Paugnat2022}, or time-dependent ones, as in Ref.~\cite{CardenasAvendano2022}. 

\subsection{Time-averaged images}

\begin{figure*}
    \centering
    \includegraphics[width=\textwidth]{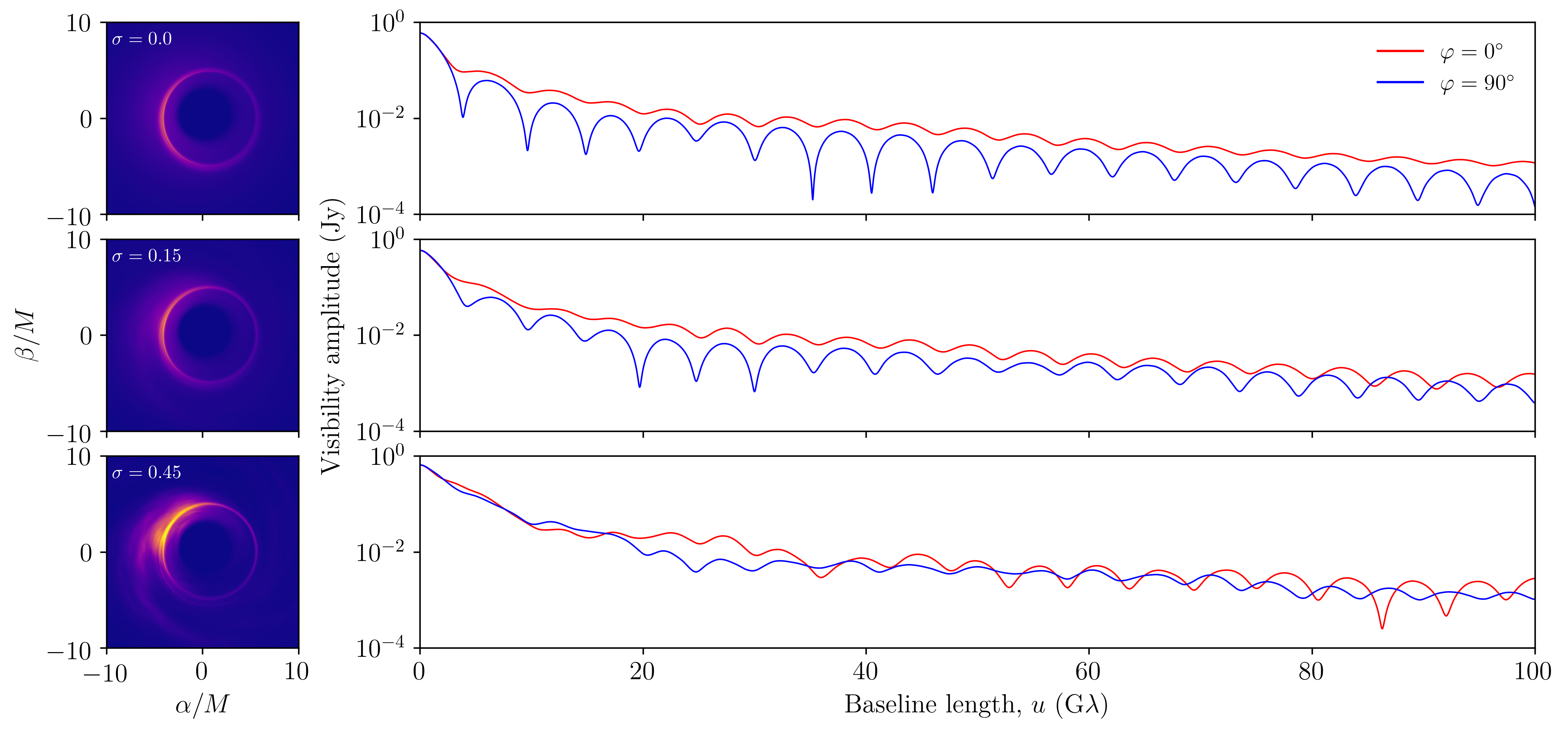}
    \caption{Black hole images (left) and their respective visibility amplitudes (right) for spin-aligned (blue) and spin-perpendicular (red) cuts across each image.
    The black hole spin is $a/M=94\%$ and the observer inclination is $\theta_{\rm o}=20^\circ$.
    From top to bottom, the magnitude $\sigma$ of the astrophysical fluctuations increases.
    In the absence of astrophysical fluctuations ($\sigma=0$), the image can be understood as being already time-averaged, with an underlying radial emission profile given by Eq.~\eqref{eq:JonhnsonSU} with the parameters of $J_{\rm SU}(r_{\rm s})$ corresponding to the profile P1 defined in Table~\ref{tbl:NominalCase}.
    These visibility amplitudes clearly show that as the amplitude $\sigma$ of the astrophysical fluctuations grows, it becomes strictly harder to identify a clear periodicity in the interferometric signal.}
    \label{fig:NoisySnapshots}
\end{figure*}

To directly obtain time-averaged images, we assume that the source emissivity is a function of radius only: $I_{\rm s}(r_{\rm s},\phi_{\rm s},t_{\rm s})=J(r_{\rm s})$.
A convenient parametric form for the radial emission profile $J(r_{\rm s})$ may be derived from the Johnson's standard unbounded (SU) distribution,
\begin{align}
	\label{eq:JonhnsonSU}
	J_{\rm SU}(r;\mu,\vartheta,\gamma)\equiv\frac{e^{-\frac{1}{2}\br{\gamma+\arcsinh\pa{\frac{r-\mu}{\vartheta}}}^2}}{\sqrt{\pa{r-\mu}^2+\vartheta^2}},
\end{align}
in which the three parameters $\mu$, $\vartheta$, and $\gamma$ respectively control the location of the profile's peak, its width, and its asymmetry. 
An example of a time-averaged image obtained in this fashion is shown in the upper-left panel of Fig.~\ref{fig:NoisySnapshots}.
Throughout this paper, we choose the location of the peak to be proportional to either the outer or inner event horizon radius, $r_{\pm}=M\pm\sqrt{M^{2}-a^{2}}$ (see Table~\ref{tbl:NominalCase}). 

\subsection{Time-dependent models}

In physical simulations, source fluctuations arise from various phenomena, including fluid turbulence, magnetic reconnection, or inherently random interactions between particles.
Here, we model all such fluctuations using the \texttt{AART} implementation~\cite{CardenasAvendano2022} of the \texttt{inoisy} code~\cite{Lee2021}, a stochastic model capable of replicating many features of realistic accretion flows onto a black hole.
This code outputs a spatiotemporally varying emissivity $I_{\rm s}(\mathbf{x}_{\rm s})$ by generating realizations of a GRF $\hat{\mathcal{F}}(\mathbf{x}_{\rm s};\mathbf{u}_i;\lambda_i)$, where the index $i\in\cu{0,1,2}$ ranges over multiple parameters: the vector $\mathbf{u}_0$ controls the temporal correlation of the flow, with characteristic correlation time $\lambda_0$, while the spatial vectors $\mathbf{u}_1$ and $\mathbf{u}_2$ determine its spatial structure, with characteristic correlation lengths $\lambda_1$ and $\lambda_2$.
The GRF is a zero-mean Mat\'ern field obtained by solving a stochastic partial differential equation \cite{Lee2021}, and it defines a source emissivity via~\cite{CardenasAvendano2022}
\begin{align}
	\label{eq:VariableSource}
	I_{\rm s}(\mathbf{x}_{\rm s})=\mathcal{J}(\mathbf{x}_{\rm s})
	\equiv J_{\rm SU}(r_{\rm s})e^{\sigma\hat{\mathcal{F}}(\mathbf{x}_{\rm s})-\frac{1}{2}\sigma^2},
\end{align}
where $\sigma$ is a parameter that controls the amplitude of the fluctuations.
In this way, the radial emission profile $J_{\rm SU}(r_{\rm s})$ becomes an envelope for the resulting stochastic source $\mathcal{J}(\mathbf{x}_{\rm s})$, and one recovers the time-averaged image when $\sigma=0$.
Examples of time-dependent snapshots are displayed in the bottom two rows of Fig.~\ref{fig:NoisySnapshots}.
In all of our simulations, we set $\lambda_0=2\pi/\Omega_{\rm K}$, where $\Omega_{\rm K}$ is the Keplerian orbital frequency, $\lambda_1=5r_{\rm s}$, and $\lambda_2=0.1\lambda_1$.
The three-dimensional unit vectors $\mathbf{u}_i$ are chosen as in Refs.~\cite{Lee2021,CardenasAvendano2022}---see, e.g., Eqs.~(109)-(111) in Ref.~\cite{CardenasAvendano2022}.
In particular, we have chosen the major axis $\mathbf{u}_1(\mathbf{x}_{\rm s})$ of the spatial correlation tensor to lie at a constant angle $\theta_\angle=20^\circ$ relative to the equatorial circles of constant radius $r_{\rm s}$, as this produces a spiral arm structure that is broadly consistent with GRMHD simulations \cite{Guan2009,Lee2021}.

\subsection{Complex visibility and visibility amplitude}

Given an image produced using Eq.~\eqref{eq:ObservedIntensity}, we compute its radio visibility as the (complex) Fourier transform
\begin{align}
	\label{eq:ComplexVisibility}
	V(\mathbf{u})=\int I_{\rm o}(\mathbf{x}_{\rm o})e^{-2\pi i\mathbf{u}\cdot\mathbf{x}_{\rm o}}\ed^2\mathbf{x}_{\rm o},
\end{align}
where $\mathbf{x}_{\rm o}=(\alpha,\beta)/r_{\rm o}$ are dimensionless coordinates (or angles) on the sky of a distant observer at large radius $r_{\rm o}\gg M$, while the sampled baseline $\mathbf{u}$ is a dimensionless distance between telescopes in the array, projected onto the plane perpendicular to the line of sight and measured in units of the observational wavelength.
As in Refs.~\cite{GLM2020,CardenasAvendano2022}, instead of taking the above $2$D Fourier transform, we apply the projection-slice theorem to compute $|V(u,\varphi)|$, the visibility amplitude along slices of fixed polar angle $\varphi$ in the Fourier plane.
To model observations of M87$^*$, we convert the Cartesian coordinates $(\alpha,\beta)$ into angles $\mathbf{x}_{\rm o}$ by using the mass-to-distance ratio $(M/r_{\rm o})_{\rm M87^*}=3.62\,\mu$as.

\noindent This effectively fixes the horizontal scale of the snapshot visibilities shown in Fig.~\ref{fig:NoisySnapshots}.
To fix their vertical scale, we normalize the visibilities such that $|V(0)|=0.6\,$Jy, which corresponds to a total compact source flux of $0.6\,$Jy \cite{Lu2023}.

In the ``universal regime'' $1/d\ll u\ll 1/w$ of baselines long enough to resolve a ring of diameter $d$, but not its width $w$, the visibility amplitude takes the universal form
\begin{align}
	\label{eq:UniversalVisibility}
    |V(u,\varphi)|&=\frac{\sqrt{\pa{\alpha_\varphi^{\rm L}}^{2}+\pa{\alpha_\varphi^{\rm R}}^{2}+2\alpha_\varphi^{\rm L}\alpha_\varphi^{\rm R}\sin\pa{2\pi d_\varphi u}}}{\sqrt{u}},
\end{align}
where $d_\varphi$ denotes the projected diameter of the ring, while the functions $\alpha_\varphi^{\rm L/R}$ encode its intensity profile \cite{Gralla2020}.

Given a simulated visibility amplitude, we attempt\footnote{As we will later see, it is not always possible to achieve this due to either added instrument noise or large astrophysical fluctuations.} to extract a projected ring diameter $d_\varphi$ at every baseline angle $\varphi$ by fitting $|V(u,\varphi)|$ to the functional form \eqref{eq:UniversalVisibility}.
Then, we can check whether the resulting $d_\varphi$ follows the GR-predicted functional form for a photon ring~\cite{GrallaLupsasca2020c,Paugnat2022},
\begin{align}
	\label{eq:Circlipse}
	\frac{d_\varphi}{2}
    &=R_{0}+\sqrt{R_1^2\sin^2\pa{\varphi-\varphi_0}+R_2^2\cos^2\pa{\varphi-\varphi_0}}.
\end{align}
This expression represents a ``circlipse'': a sum of a circle (with radius $R_0$) and an ellipse (with axes $R_1$ and $R_2$).
The phase $\varphi_0$ adds some flexibility to accommodate for rotations of the image.
To extract the diameters $d_\varphi$ from Eq.~\eqref{eq:UniversalVisibility}, we use the fitting method described in the next section, which is a version of the procedure presented in Ref.~\cite{Paugnat2022}, but generalized to take into account the effects of instrument noise and astrophysical fluctuations.

\subsection{Fitting method}
\label{subsec:FittingMethod}

Given any profile, we consider $36$ visibility amplitudes computed along baseline angles $\varphi\in\cu{0^\circ,5^\circ,\ldots,175^\circ}$, where the angle $\varphi\in[0, 2\pi)$ also parameterizes $d_\varphi$ and $\alpha_\varphi^{\rm L/R}$ \cite{GrallaLupsasca2020c}.
For each baseline angle $\varphi$, we use \texttt{SciPy}'s simple fitting routine \verb|curve_fit| to fit the sampled points of the visibility amplitude $|V(u,\varphi)|$ to the functional form \eqref{eq:UniversalVisibility}, resulting in some best-fit parameters $d_\varphi$ and $\alpha_\varphi^{\rm L/R}$.

For this fitting method to work in practice, we must provide suitable bounds and initial guesses for the three fitting parameters.
To set approximate bounds for $\alpha_\varphi^{\rm L/R}$, we use Eq.~\eqref{eq:UniversalVisibility} to derive relations between their extrema and those of the visibility amplitude $|V(u,\varphi)|$ on the cut:
\begin{align}
    \alpha_{\rm max}^{\rm L*}+\alpha_{\rm max}^{\rm R*}&\approx V_{\rm max}\sqrt{u_{\rm max}}, \\
    \alpha_{\rm max}^{\rm L*}-\alpha_{\rm max}^{\rm R*}&\approx V_{\rm min}\sqrt{u_{\rm min}}.
\end{align}
Solving these equations yields the approximate bounds
\begin{align}
    \alpha_{\rm max}^{\rm L/R}&=\frac{1}{2}\pa{V_{\rm max}\sqrt{u_{\rm max}}\pm V_{\rm min}\sqrt{u_{\rm min}}}.
\end{align}
These vary across the different baseline angles $\varphi$, since the visibility profile along each cut is generally different.
As for the lower bounds on $\alpha_\varphi^{\rm L/R}$, we set them to zero. 

For the diameter, we impose the prior $d_\varphi\in[35,45]\,\mu$as, since our analysis is based on M87$^*$, and photon ring diameters outside this range are excluded on physical grounds.
The initial guess for $d_\varphi$ is always set to $40\,\mu$as and the initial guess for $\alpha_\varphi^{\rm L/R}$ to $k\cdot\alpha_{\rm max}^{\rm L/R}$, where $k$ is some positive constant less than unity.
To fix a value for $k$, we tried several fits with $k\in\cu{0.05, 0.10,\ldots,0.95}$ and recorded the average root-mean-square deviation (RMSD) for each trial, as defined in Eq.~\eqref{eq:RMSD} below.
We found that a value of $k=0.15$ consistently gave good fits, that is, it consistently led to the lowest RMSD.

Having obtained best-fit parameters $d_\varphi$ and $\alpha_\varphi^{\rm L/R}$ at every angle $\varphi$, the subsequent parts of the fitting method closely follow the procedure described in Ref.~\cite{Paugnat2022}.
That is, we proceed by surveying the multi-peaked distribution of the RMSD, defined by an average $\av{}_u$ over baselines as
\begin{align}
	\label{eq:RMSD}
    \text{RMSD}_u(d)=\frac{\sqrt{\av{\br{|V_{\rm fit}(u;d)|-|V(u)|}^{2}}_u}}{\av{|V_{\rm fit}(u;d)|}_u}.
\end{align}
However, instead of considering an adaptive range around $d_\varphi$ with a width determined by the power associated with the baseline window, as in Ref.~\cite{Paugnat2022}, we instead fix the range of candidate ring diameters to $[8M,12M]$.
Again, for M87* profiles, this is justifiable on physical grounds.

Back in line with Ref.~\cite{Paugnat2022}, we then survey diameters $d_\varphi$ that maximize the goodness-of-fit $g(d_\varphi)$, defined as
\begin{align}
    g(d_\varphi)\equiv e^{-\text{RMSD}_u(d_\varphi)},
\end{align}
and retain those for which $g(d_\varphi)$ exceeds some threshold, which, throughout our analysis, we set to be $g(d_\varphi)\ge 0.1$.

These candidate angle-dependent diameters $d_\varphi$ form multiple distinct circlipses, which we denote by $\mathcal{C}_i$.
The $\mathcal{C}_i$ are separated by a gap $\Delta\mathcal{C}_i\approx 1/u_w$, where $u_w$ is the length of the baseline in the middle of the chosen baseline window.
Each $\mathcal{C}_i$ has a joint goodness-of-fit given by 
\begin{align}
	\label{eq:JointGoodness}
	g(\mathcal{C}_i)&=\prod_{\varphi=0^\circ}^{175^\circ}g(d_\varphi).
\end{align}
The most probable circlipse is the one with the largest joint goodness-of-fit $g(\mathcal{C}_i)$.
We fit it to the GR-predicted functional form \eqref{eq:Circlipse} for the photon ring diameter to find the parameters $\mathbf{R}=\cu{R_0,R_1,R_2,\varphi_0}$ that minimize
\begin{align}
    \text{RMSD}_\varphi(\mathbf{R})=\frac{\sqrt{\av{\br{d_{\rm fit}(\varphi;\mathbf{R})-d_\varphi}^2}_\varphi}}{\av{d_{\rm fit}(\varphi;\mathbf{R})}_\varphi}.
\end{align}

For various reasons, one must select a narrow baseline window when performing a fit~\cite{Paugnat2022}, and the characteristic baseline length $u_w$ of the window is especially important.
Indeed, $u_w$ determines the specific photon ring whose geometry the fits are probing, since different subrings dominate different regions of the visibility domain \cite{JohnsonLupsasca2020}.

In particular, a choice of baseline window that falls within a transition region between regimes dominated by one photon subring and the next is less favorable.
In such regions, interference between the signals of the two rings prevents the emergence of a ``clear'' interferometric signature~\cite{CardenasAvendano2023}, thereby inhibiting our ability to recover information about the shape of either photon ring.

Since we are particularly interested in measurements of the $n=1$ photon ring, we consider baseline windows in the relatively shorter range between $u\in[0,150]\,\text{G}\lambda$.\footnote{For example, for observations at $345$\,GHz, a baseline of $150\,\text{G}\lambda$ translates to a telescope separation of about 10 Earth diameters.}

This concludes our presentation of the elements needed to investigate measurements of the first photon ring in the presence of astrophysical fluctuations and instrument noise.
Before proceeding with our analysis, we briefly summarize the steps in our methodology.
First, we ray trace simulated black hole images using Eq.~\eqref{eq:ObservedIntensity}, in either a time-averaged model with no astrophysical fluctuations or a time-dependent one.
For a given image, we compute its visibility amplitude at various baseline angles $\varphi$ via Eq.~\eqref{eq:ComplexVisibility}.
Then, we use the fitting method outlined in Sec.~\ref{subsec:FittingMethod} to extract a ring shape $d_\varphi$ from the visibility amplitude, which we compare to the GR prediction \eqref{eq:Circlipse}.

In the next sections, we study each effect separately, starting with the impact of instrument noise, followed by that of fluctuations, and finally, an example with both.

\section{The impact of instrument noise}
\label{sec:InstrumentNoise}

In this section, we investigate the effect of instrument noise on the inferred shape of the first photon ring.
In particular, we consider how instrument noise affects the accuracy and precision of our fitting procedure, and how much noise our method can handle before breaking down.
After finding its limits, we close this section by discussing which choice of baseline window might be most conducive to accurate measurements of the first photon ring.

Given a simulated complex visibility \eqref{eq:ComplexVisibility}, we model the effects of instrument noise by adding to it a realization of a complex Gaussian distribution $\mathcal{N}_{\mathcal{C}}(0,s)$ with mean zero and standard deviation $s$.

After we add instrument noise, we would like to know how much of an effect it has on the underlying signal.
To understand this, it is instructive to keep track of the power associated with a given visibility amplitude.
For any given $|V(u)|$, we define the power of the visibility amplitude across a baseline window $[u_1,u_2]$ as
\begin{align}
    \label{eq:Power}
    P_\varphi(u_1,u_2)=\frac{\ab{V(u_1,\varphi)}+\ab{V(u_2,\varphi)}}{2}.
\end{align}

This power is a measure of the strength of the visibility amplitude in the baseline range $[u_1,u_2]$ at fixed angle $\varphi$.
Of course, it varies with the overall normalization of the visibility amplitude.
This notion of power is especially important when thinking about the significance of some level of noise, since it provides---in combination with $s$---a rough estimate of the signal-to-noise ratio (SNR).

\subsection{Accuracy and precision of the inferred interferometric diameters}

We begin by assessing the effect of instrument noise on the accuracy and precision with which we can infer an interferometric ring diameter using our fitting method.

We consider a Kerr black hole with spin $a/M=94\%$, observed from an inclination of $\theta_{\rm o}=20^\circ$.
We choose the time-averaged emission profile to be given by Eq.~\eqref{eq:JonhnsonSU} with parameters $\mu=r_-$, $\gamma=-3M/2$, and $\vartheta=M/2$.
This profile is listed as P1 in Table~\ref{tbl:NominalCase}.

\begin{table}
    \centering
    \begin{tabular}{cc}
    \hline 
    Parameter & Value \tabularnewline
    \hline 
    \hline 
    $a/M$ & $94\%$ \tabularnewline
    $\theta_{\rm o}$ & $20^\circ$ \tabularnewline
    P$1=(\mu,\vartheta/M,\gamma)$ & $\pa{r_-,\frac{1}{2},-\frac{3}{2}}$ \tabularnewline
    P$2=(\mu,\vartheta/M,\gamma)$ & $\pa{\frac{3}{2}\,r_+,1,0}$
    \tabularnewline
    \hline 
    \hline 
    \end{tabular}
    \caption{The black hole spin and inclination of the example case we consider throughout this paper, and the parameters for our two radial emission profiles \eqref{eq:JonhnsonSU}.
    The outer/inner event horizon radii are denoted by $r_{\pm}=M\pm\sqrt{M^2-a^2}$.}
    \label{tbl:NominalCase}
\end{table}

We carry out $500$ fits with distinct noise realizations, first with a noise level of $s=1.5\,$mJy and then again with $s=2.5\,$mJy, across the baseline window $[86,116]\,\text{G}\lambda$.\footnote{For a noise level of $s=1.5\,$mJy, Fig.~\ref{fig:BaselineWindowSize} shows that this baseline window gives the lowest average RMSD over all the profiles we study in this work (see Sec.~\ref{subsec:SuitableWindows} for details).
Furthermore, as we will see shortly, inferring an interferometric diameter across this particular baseline window is quite challenging when using the astrophysical parameters of the profile P1 listed in Table~\ref{tbl:NominalCase}.}

Over the $36$ baseline angles $\varphi$ that we examine within this window, the visibility amplitude attains a minimum power \eqref{eq:Power} of $0.56\,$mJy, a maximum value of $1.18\,$mJy, and an average value of $0.84\,$mJy.
As shown in the right panels of Fig.~\ref{fig:MultiPeakedCirclipses}, the instrument noise levels $s=1.5\,$mJy and $s=2.5\,$mJy really do affect the visibility amplitude and its underlying interferometric signal.

For each successful fit, we compute the maximal (and orthogonal) ring diameters
\begin{align}
	d_\parallel&=2(R_0+R_1),\\
    d_\perp&=2(R_0+R_2),    
\end{align}
where $R_0$ and $R_2$ are the best-fit parameters in Eq.~\eqref{eq:Circlipse}.
When $\varphi_0=0$, these are exactly the diameters parallel and perpendicular to the projected black hole spin axis.

Together, they define the fractional ring asymmetry
\begin{align}
	\label{eq:FractionalAsymmetry}
    f_A=1-\frac{d_\perp}{d_\parallel},
\end{align}
which characterizes the ``shape'' of the circlipse.

\begin{figure}
    \centering
    \includegraphics[width=\columnwidth]{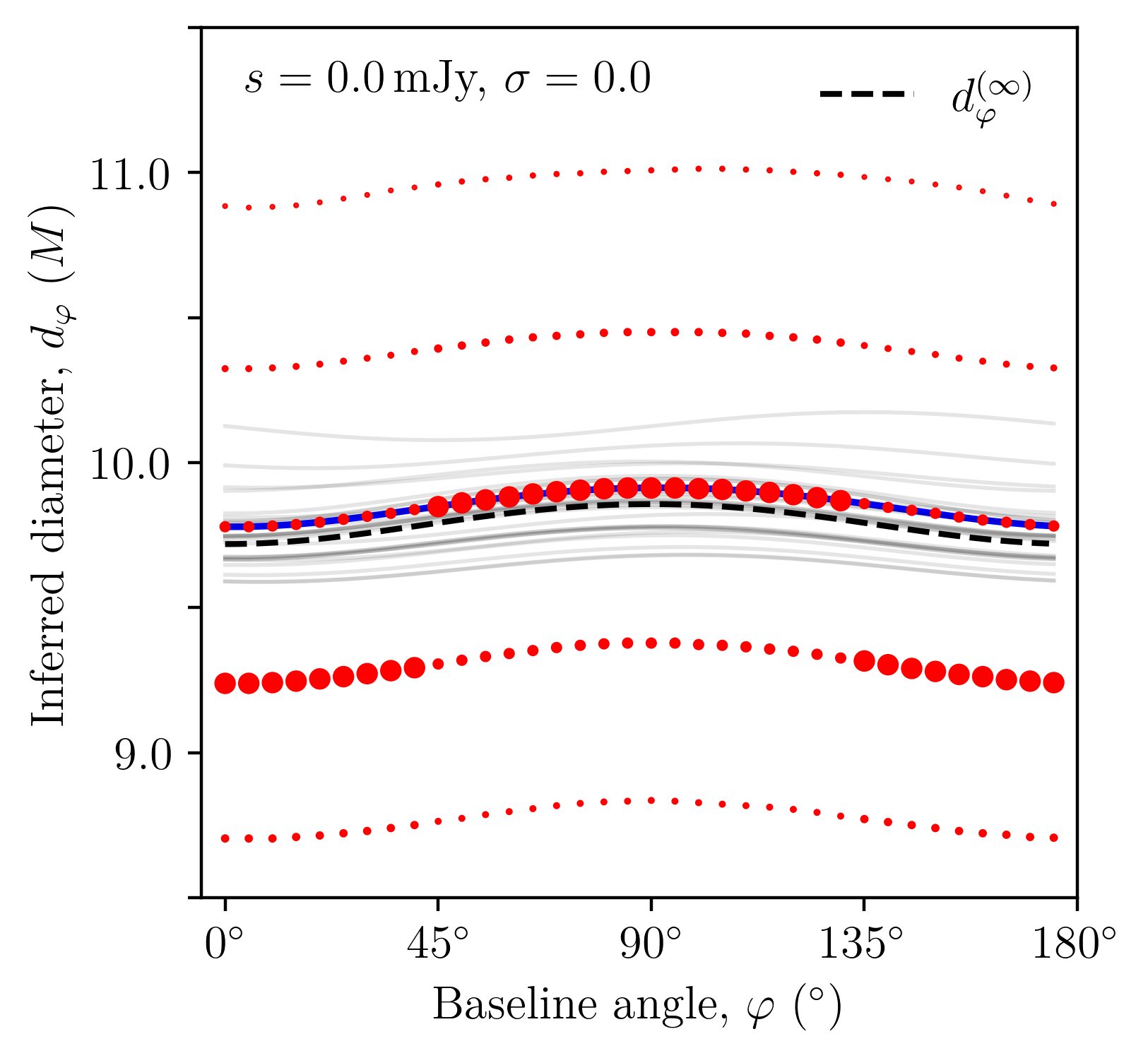}
    \caption{The multi-peaked distribution of inferred photon ring diameters in the absence of noise or astrophysical fluctuations, for a black hole with spin $a/M=94\%$ surrounded by the emission profile P1 specified in Table~\ref{tbl:NominalCase} and observed from an inclination of $\theta_{\rm o}=20^\circ$.
    For each baseline angle $\varphi$, we plot the possible inferred diameters with dots whose sizes are set by the relative magnitudes of their goodness-of-fit (and not their absolute values).
    That is, the point sizes should only be compared vertically (i.e., at fixed $\varphi$) in order to identify which diameter had the highest goodness-of-fit, the second highest, and so on, at that particular baseline angle.
    The dashed black line is the circlipse that best fits the critical curve, while the background shaded grey lines are the circlipses inferred from the same black hole, but with the $30$ different astrophysical profiles specified by the parameters in Table~\ref{tbl:ProfileParameters}.
    This figure and the upper-left panel of Fig.~\ref{fig:MultiPeakedCirclipses} both display the same data, including the same inferred circlipse (blue curve).
    However, here we use point sizes rather than colors to better illustrate the differences in the goodness-of-fits of the inferred diameters and to highlight the ``jumps'' in the best-fit diameters that occur at $d_\varphi\approx9.2M$ and $d_\varphi\approx9.8M$.}
\label{fig:CirclipseInference}
\end{figure}

\begin{figure*}
    \centering
    \includegraphics[width=\textwidth]{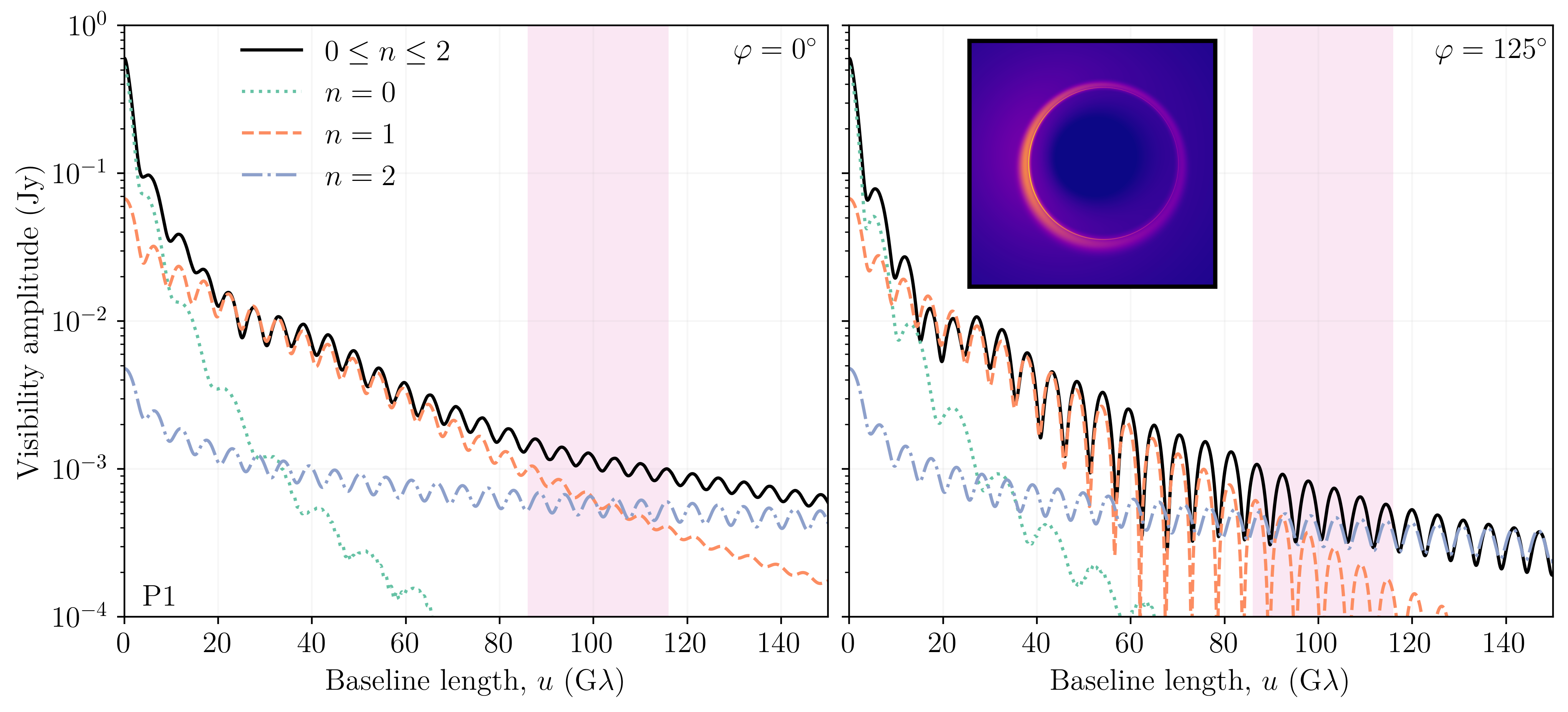}
    \caption{The visibility amplitude corresponding to the direct image ($n=0$, green-dotted lines), the first two photon rings ($n=1$, orange-dashed lines, and $n=2$, blue-dash-dotted lines), and the total image ($0\leq n\leq2$, black solid lines) for the profile P1 (see Table~\ref{tbl:NominalCase}), in the absence of instrument noise and astrophysical fluctuations.
    The pink-shaded region corresponds to the baseline window $[86,116]\,\text{G}\lambda$ over which we attempt to infer an interferometric diameter.
    The left and right panels display the visibility amplitude at baseline angles $\varphi=0^\circ$ and $\varphi=125^\circ$, respectively.
    The right panel inset shows the corresponding image.
    In this baseline window, the visibility undergoes at every angle $\varphi$ a transition between regimes dominated by the $n=1$ and $n=2$ photon rings.
    Nevertheless, one can still infer a circlipse shape for $d_\varphi$, as shown in Figs.~\ref{fig:CirclipseInference} and \ref{fig:MultiPeakedCirclipses} (upper-left panel).}
    \label{fig:Transitions}
\end{figure*}

As described in Sec.~\ref{subsec:FittingMethod}, for each fitting attempt, we generate multiple circlipses $\mathcal{C}_i$ by surveying the multi-peaked distribution of the RMSD \eqref{eq:RMSD}, and then we fit the circlipse with the highest joint goodness-of-fit \eqref{eq:JointGoodness} to the expected functional form \eqref{eq:Circlipse} for the photon ring shape in order to infer its angle-dependent diameter $d_\varphi$.

We draw the best-fit circlipse in the absence of noise and fluctuations as a blue line in Fig.~\ref{fig:CirclipseInference} and the upper-left panel of Fig.~\ref{fig:MultiPeakedCirclipses}.
The first shows the relative distribution of the goodness-of-fits for the inferred diameters, while the second offers a more faithful representation across angles.
This facilitates a comparison of the joint goodness-of-fits across all the candidate circlipses.

We see in Fig.~\ref{fig:CirclipseInference} that for the emission profile P1 given in Table~\ref{tbl:NominalCase}, which is broadly consistent with the EHT results of M87$^*$ on Earth baselines~\cite{GLM2020}, the inferred diameters $d_\varphi$ with the highest goodness-of-fit can ``jump'' between two circlipses at certain angles $\varphi$.
These discontinuities arise as a result of the choice of baseline window, and their presence and location depend on the set of astrophysical parameters under consideration.
In the chosen window, the visibility amplitude of profile P1 is \emph{always} sampled (at every angle $\varphi$) in the transition region between the regimes dominated by the first and second photon rings (see Fig.~\ref{fig:Transitions}).
That is, for all of the $36$ sampled baseline angles, the first photon ring never dominates the signal across the entire baseline window in which we perform the fits.
As discussed in detail in Ref.~\cite{Paugnat2022}, a diameter inferred in such a setting lacks a very sharp geometric interpretation: if neither the $n=1$ nor the $n=2$ photon ring dominates the visibility amplitude, then its ringing is not directly related to the physical size of either ring; rather, it is a feature of the continuity of the signal and its periodicity~\cite{Paugnat2022}.
In other words, when sampling in a transition region, one can still infer an interferometric diameter $d_\varphi$ by fitting the universal form \eqref{eq:UniversalVisibility}, but this diameter lacks a clear interpretation in the image.

Despite this jump in the inferred diameter $d_\varphi$, one can still (in some cases) infer a ``correct'' circlipse shape, as shown for instance in the lower panel of Fig.~4 of Ref.~\cite{CardenasAvendano2023}.
(The panel shows that a ``good'' circlipse can be inferred in the baseline window $[70,100]\,\text{G}\lambda$, but not the window $[40,70]\,\text{G}\lambda$, for the emission profile P2 given in Table~\ref{tbl:NominalCase}). 
This is not too surprising because the visibility amplitude is dominated by a ring on either side of the transition region, so its ringing encodes a circlipse shape on both sides of the transition: by continuity, we are then led to expect the interferometric diameter within the transition region to also follow a circlipse shape.

Indeed, the circlipse shape we infer in this case is still the ``correct'' one, that is, the one closest to the circlipse shape associated with the critical curve (the dashed black line in Fig.~\ref{fig:CirclipseInference}).
As we vary the underlying astrophysical profile $J_{\rm SU}(r_{\rm s})$ across the set of $30$ parameters specified in Table~\ref{tbl:ProfileParameters}, we continue to infer similar circlipse shapes that cluster in a band around the same location (see the shaded grey background lines in Fig.~\ref{fig:CirclipseInference}).

\begin{figure*}
    \centering
    \includegraphics[width=\textwidth]{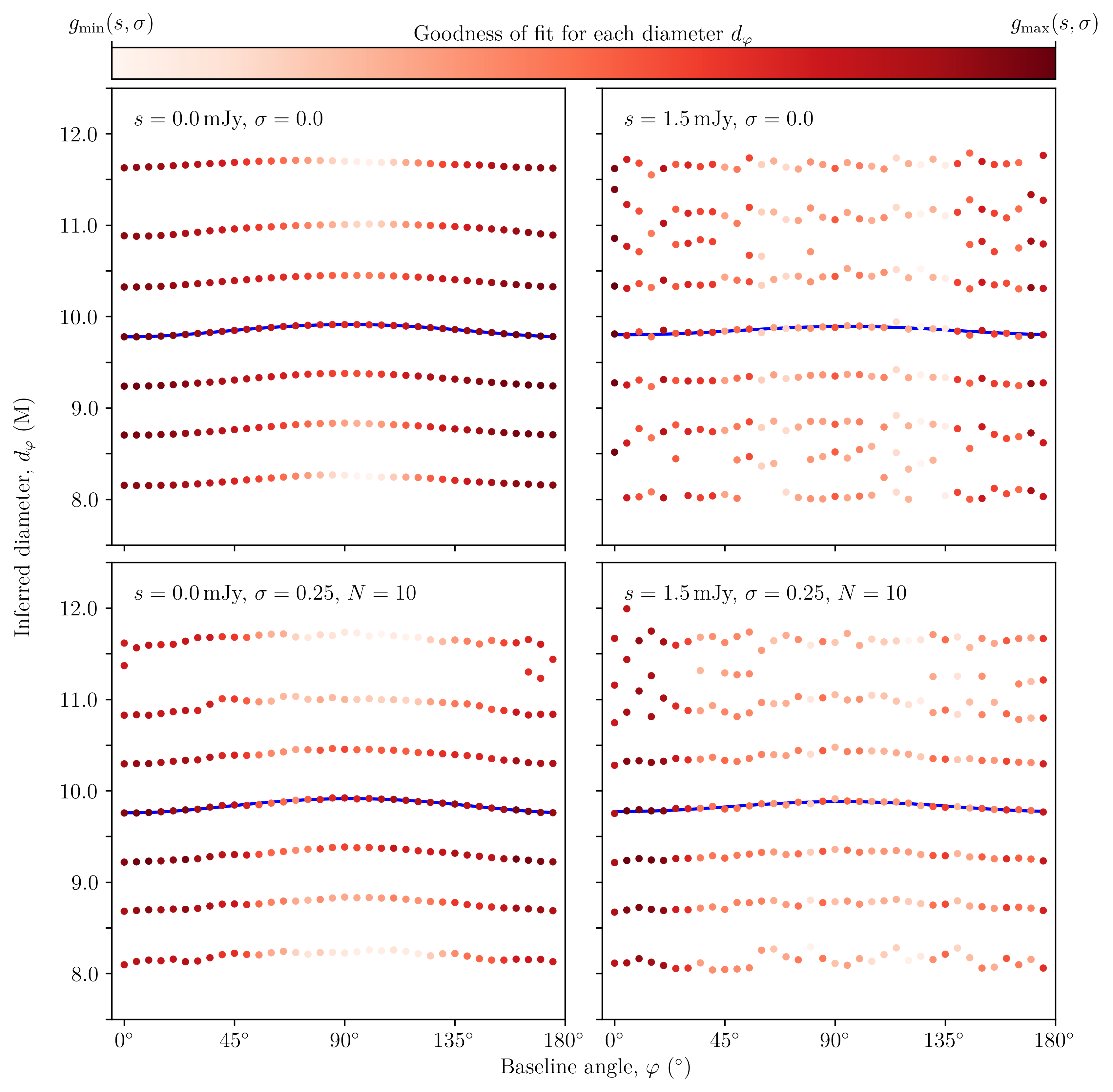}
    \caption{The multi-peaked distribution of inferred photon ring diameters for a black hole with spin $a/M=94\%$ surrounded by the emission profile P1 specified in Table~\ref{tbl:NominalCase} and observed from an inclination of $\theta_{\rm o}=20^\circ$.
    The four panels display simulated measurements with various combinations of instrument noise and astrophysical fluctuations across the window $[86,116]\,\text{G}\lambda$: no noise or fluctuations (top left), $s=1.5\,$mJy of instrument noise alone (top right), astrophysical fluctuations with $\sigma=0.25$ alone (bottom left), and a combination of both instrument noise and astrophysical fluctuations with $s=1.5\,$mJy and $\sigma=0.25$, respectively (bottom right).
    In the absence of astrophysical fluctuations ($\sigma=0$), realizations of the complex Gaussian noise $\mathcal{N}_{\mathcal{C}}(0, s)$ are added to the complex visibility across each baseline angle $\varphi$.
    In the presence of fluctuations ($\sigma>0$), we carry out the same process separately across $N=10$ snapshots, before taking their coherent time average.
    In either case, we then fit the sampled visibility amplitude across all angles to the GR-predicted functional form \eqref{eq:UniversalVisibility} to obtain the ring diameter $d_\varphi$ that locally minimizes the RMSD \eqref{eq:RMSD}.
    We also consider as possible diameters the other local minima of the RMSD (or, equivalently, the local maxima of the goodness-of-fit) in the range $[8M,12M]$.
    Each panel has its own color scale, whose limits are set by the absolute maximum and minimum of the goodness-of-fit for all the inferred diameters across all baseline angles.
    The multi-peaked diameter distribution gives rise to multiple circlipse shapes, and the blue lines correspond to the circlipse with the highest joint goodness-of-fit \eqref{eq:JointGoodness} in each fitting attempt.}
	\label{fig:MultiPeakedCirclipses}
\end{figure*}

\begin{figure*}
    \centering
    \includegraphics[width=0.98\textwidth]{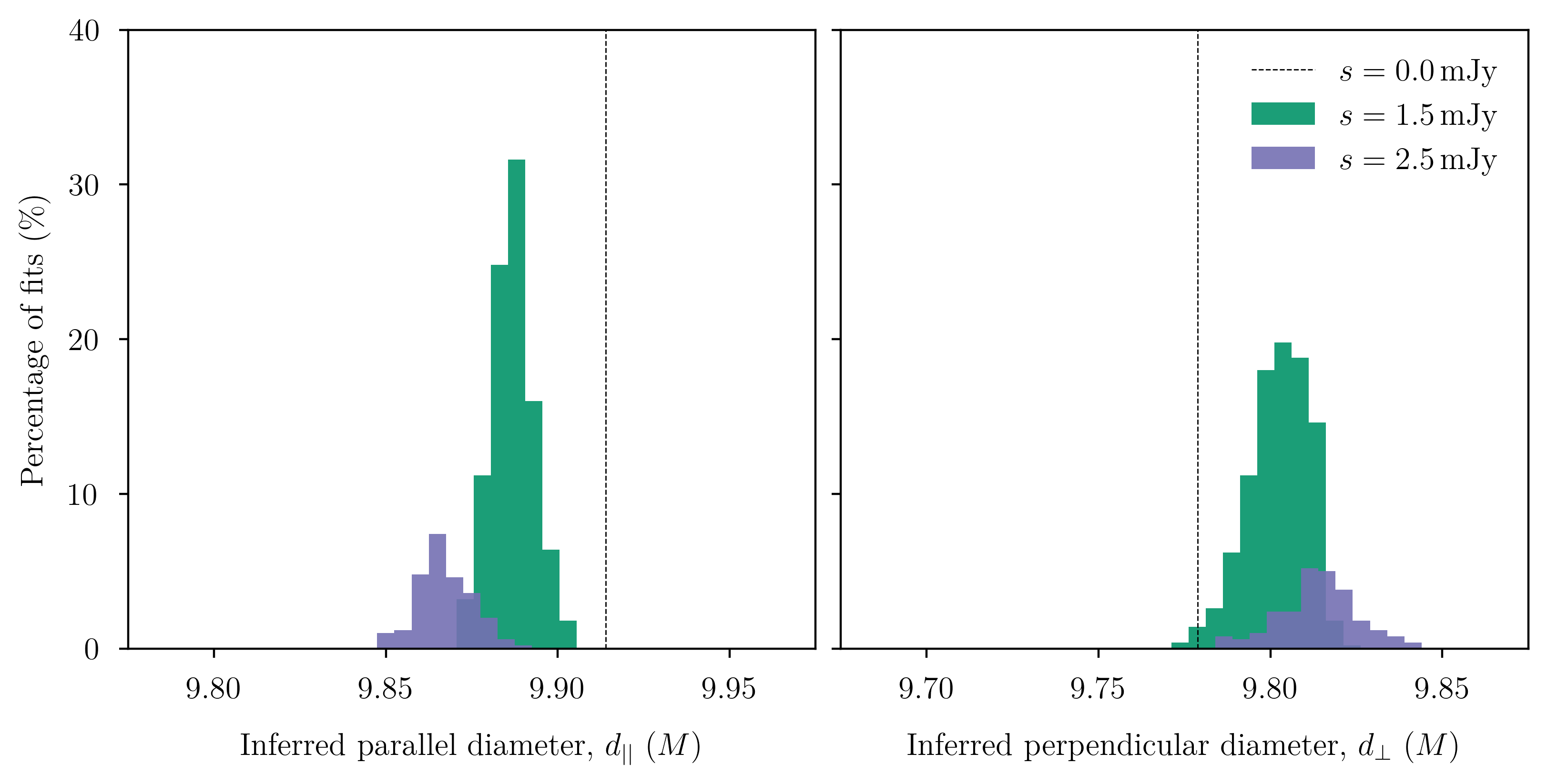}
    \caption{The inferred-diameter distributions for the parallel diameter $d_\parallel$ (left) and the perpendicular diameter $d_\perp$ (right), for a black hole with spin $a/M=94\%$ surrounded by the emission profile P1 specified in Table~\ref{tbl:NominalCase} and observed from an inclination of $\theta_{\rm o}=20^\circ$.
    The fits are performed in the baseline window $[86,116]\,\text{G}\lambda$ with $500$ realizations of instrument noise for $s=1.5\,$mJy (green histograms) and another $500$ for $s=2.5\,$mJy (purple histograms).
    The presence of instrument noise shifts the centers of the distributions relative to the diameters inferred in the absence of noise (i.e., $s=0\,$mJy), denoted by the vertical dashed lines.
    For the inferred parallel diameters, the centers of the $s=1.5\,$mJy and $s=2.5\,$mJy histograms are shifted by $0.25\%$ and $0.46\%$, respectively, relative to the inferred $d_\parallel$ in the absence of noise.
    Similar shifts, but in the opposite direction, occur for the inferred $d_\perp$ histograms.
    As expected, as the level of noise increases, the widths of the histograms also increase.
    In particular, as the noise level grows from $s=1.5\,$mJy to $s=2.5\,$mJy, the widths of the $d_\parallel$ and $d_\perp$ histograms increase by factors of $\approx1.3$, and $\approx1.1$, respectively.
    The most pronounced effect of increased instrument noise is a higher rate of fit failure (i.e., a decrease in the area of the histograms): $95\%$ of the fits are successful when $s=1.5\,$mJy, while only $25\%$ of fits succeed when $s=2.5\,$mJy.}
\label{fig:InferredDiameterDistributions}
\end{figure*}

As shown in the upper-right panel of Fig.~\ref{fig:MultiPeakedCirclipses}, we are still able to infer a circlipse in the correct band, even using noisy observations in a transition region.
This circlipse, however, is slightly distorted and displaced relative to the one inferred in the absence of noise.
Since each of the $500$ fitting attempts in this survey has a different realization of noise, we expect the circlipses inferred from each fit to be deformed in slightly different ways.
Indeed, the different diameters $d_\varphi$ inferred from each fit produce the distribution shown in Fig.~\ref{fig:InferredDiameterDistributions}.
For example, the inferred parallel diameter $d_\parallel$ is given, for each realization of noise, by the ordinates of the blue curves at $\varphi=90^\circ$ in Fig.~\ref{fig:MultiPeakedCirclipses}.

For each realization of noise, the inferred circlipse is slightly displaced vertically, leading to small differences in the inferred values of $d_\parallel$ across the $500$ fits. 

The distribution of diameters shown in Fig.~\ref{fig:InferredDiameterDistributions} is to be distinguished from the multi-peaked distribution of diameters presented in Fig.~\ref{fig:MultiPeakedCirclipses}; the latter is due to the RMSD having various local minima, each of which may or may not correspond to the global minimum, while the former is due to different realizations of instrument noise, each of which leads to an inferred circlipse that is the same as in the noiseless ($s=0\,$mJy) case, but slightly displaced and distorted by the effects of noise. 

To assess the accuracy of these diameter distributions, we use as a benchmark the diameters $d_\parallel$ and $d_\perp$ inferred in the noiseless case ($s=0\,$mJy), which are plotted as dashed vertical lines in Fig.~\ref{fig:InferredDiameterDistributions}.
Although these diameters do not correspond to the interferometric diameters of the $n=1$ photon ring (since the visibility amplitude is in a region of transition at all baseline angles), nonetheless they still offer insight into the accuracy of our diameter inference.
This is because, as Fig.~\ref{fig:CirclipseInference} shows, the circlipse they correspond to is remarkably close to the shape of the critical curve.
In particular, the diameters $d_\parallel$ and $d_\perp$ inferred in the absence of noise differ from those of the critical curve by $0.58\%$ and $0.61\%$, respectively.

Moreover, the RMSD of the fit to the functional form of the circlipse is $0.008\%$ in the absence of noise.
The close proximity of the inferred circlipse to the critical curve shape, together with the low RMSD of the fit, suggest that, at least in the absence of noise and astrophysical fluctuations, our fitting scheme is capable of recovering information about the underlying geometry, even from a transition region.
As such, we can use these diameters as a benchmark to assess the accuracy of the distributions of diameters inferred in the presence of instrumental noise (and astrophysical fluctuations).

\begin{figure*}
    \centering
    \includegraphics[width=0.95\textwidth]{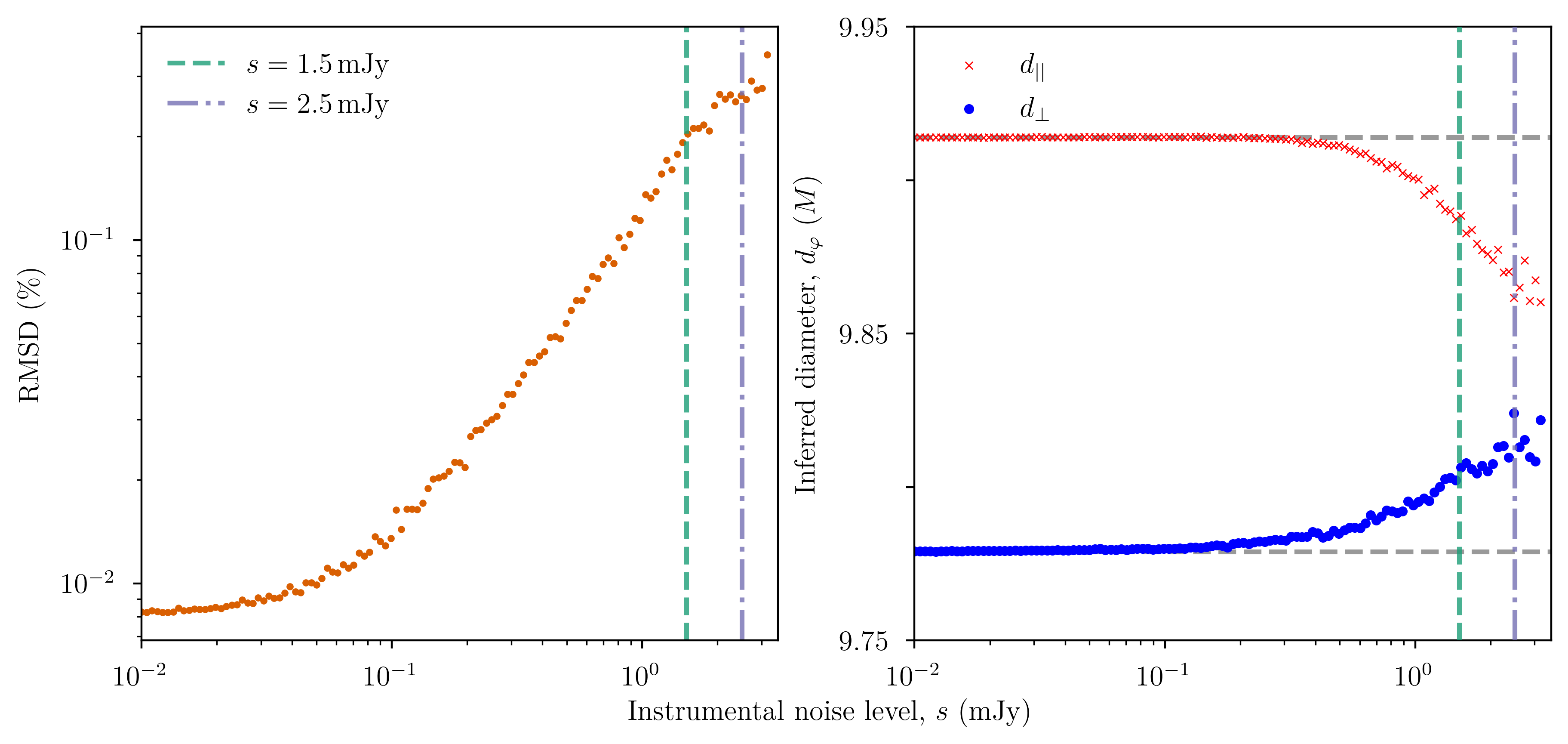}
    \caption{Average RMSD values (left) and inferred diameters (right) for $15$ fitting attempts across the baseline window $[86,116]\,\text{G}\lambda$ in the presence of instrument noise ranging from a level $s=10^{-5}\,$Jy to $s=10^0\,$Jy.
    The vertical lines indicate the noise levels $s=1.5\,$mJy (green dashed line) and $s=2.5\,$mJy (purple dash-dotted line), and the horizontal lines on the right panel indicate the diameters inferred in the absence of noise using the envelope-fitting method of Ref.~\cite{Paugnat2022}.
    At the largest plotted noise level of $s=3.2\,$mJy, the average inferred diameters $d_\parallel$ (red crosses) and $d_\perp$ (blue dots) differ by $0.54\%$ and $0.44\%$, respectively, from those inferred in the absence of noise.
    For noise levels above $s=3.2\,$mJy, the RMSD plateaus and successful fits become much less frequent: for $s\leq 3.2\,$mJy, $93\%$ of all the fitting attempts succeed, whereas only $1\%$ succeed for $s>3.2\,$mJy.}
    \label{fig:NoiseThreshold}
\end{figure*}

As Fig.~\ref{fig:InferredDiameterDistributions} shows, adding noise causes the mean of each histogram to shift slightly relative to the dashed vertical lines showing the diameters inferred in the absence of instrument noise.
In particular, for $d_\parallel$, the means of the $s=1.5\,$mJy and $s=2.5\,$mJy histograms are displaced by $0.25\%$ and $0.46\%$, respectively, relative to the value of $d_\parallel$ inferred in the absence of noise, with shifts of comparable size for $d_\perp$.
Moreover, the width of the histograms for $d_\parallel$ (or $d_\perp$) increases by a factor of $\approx1.3$ (or $\approx1.1$) as the noise level increases from $s=1.5\,$mJy to $s=2.5\,$mJy. 

The changes we observe in the mean values and widths of the histograms in Fig.~\ref{fig:InferredDiameterDistributions} are quite modest relative to the changes in their areas.
In other words, the strongest effect of increased noise levels is an increased failure rate: $95\%$ of the fits are successful when this level is $s=1.5\,$mJy, but that figure decreases to $25\%$ when the noise grows to $s=2.5\,$mJy.
As such, increasing the level of instrument noise leads to a significant reduction in the likelihood of a successful fit; on the other hand, for successful fits, increasing the noise level only leads to a modest reduction in the accuracy and precision of the fit.
In sum, the measurements of $d_\varphi$ are either good, or fail.

Since Fig.~\ref{fig:InferredDiameterDistributions} includes the full range of diameters $d_\varphi$ inferred from our survey, we see that increasing the level of instrument noise never leads us (in any of the $500$ fits) to select the second circlipse, which exhibits jumps in $d_\varphi$.

\subsection{Breakdown of our fitting method}

Having assessed the accuracy and precision of our fits in the presence of a few levels of instrument noise, we now vary the noise level $s$ to find out how much of it our fitting method can handle before breaking down.

We perform fits to the visibility amplitude of a black hole with profile P1 (defined in Table~\ref{tbl:NominalCase}) in the baseline window $[86,116]\,\text{G}\lambda$, in the presence of $15$ realizations of noise and for $250$ different noise levels chosen to grow exponentially from $s=10^{-5}\,$Jy to $s=1\,$Jy.
The results are shown in Fig.~\ref{fig:NoiseThreshold}, wherein the ordinate of each point in the left (or right) plot is the average RMSD (or inferred diameter) of the $15$ fits attempted at the corresponding level of noise.
The dashed vertical lines denote the noise levels $s=1.5\,$mJy and $s=2.5\,$mJy from the last section.

The last points plotted in Fig.~\ref{fig:NoiseThreshold} correspond to a noise level $s_{\rm max}=3.2\,$mJy, beyond which fewer than three successful fits can be obtained for any given $s$.
Indeed, the overwhelming majority of fitting attempts above this level of noise fail:\footnote{The very last successful fit we obtained had a noise of $s=0.53\,$Jy.
The diameters $d_\parallel$ and $d_\perp$ that we inferred differed by $0.54\%$ and $0.28\%$, respectively, from those inferred in the absence of noise.}
for $s\leq s_{\rm max}$, $93\%$ of the fitting attempts are successful, whereas only $1\%$ are successful for $s>s_{\rm max}$.
The particular level of noise $s=3.2\,$mJy is only significant insofar as it relates to the power of the visibility amplitude across $[86,116]\,\text{G}\lambda$: the ratios of $s_{\rm max}$ to the maximum and average powers are $s_{\rm max}/P_{\rm max}=2.7$ and $s_{\rm max}/P_{\rm avg}=3.8$.
In other words, the success rate of the fits drops below $20\%$ when the noise level $s$ exceeds $3.8$ times the (average) power.

As the left panel of Fig.~\ref{fig:NoiseThreshold} shows, the RMSD starts to grow significantly around $s=0.1\,$mJy.
By contrast, the inferred diameters remain relatively insensitive to this depreciation of the quality of fit (as suggested by a higher RMSD).
Indeed, the percentage change in the diameters remains modest over the entire range of noise levels in Fig.~\ref{fig:NoiseThreshold}.
In particular, for $s=3.2\,$mJy (the highest noise value plotted), the average inferred $d_\parallel$ and $d_\perp$ differ from their fitted values in the absence of noise by only $0.54\%$ and $0.44\%$, respectively. 

This small variation of the inferred diameters $d_\varphi$ with respect to the noise level explains the shifts observed in the histogram peaks in Fig.~\ref{fig:InferredDiameterDistributions}.
For the largest values of $s$ in Fig.~\ref{fig:NoiseThreshold}, the inferred diameters asymptote to a single value.
For even larger values of $s$ beyond those shown in Fig.~\ref{fig:NoiseThreshold}, the inferred shape becomes more circular, which corresponds to the flattening of the inferred circlipse.

This can be visually understood from the upper-right panel of Fig.~\ref{fig:MultiPeakedCirclipses}: when there is a significant increase in the noise, the inferred diameters start to oscillate around a common value.
Consequently, the ring shape and its asymmetry get washed out, and one can only infer a straight line corresponding to a circular radius. 

Remarkably, \emph{all} the diameters inferred in this survey lie within the range delineated by the two horizontal dashed lines in Fig.~\ref{fig:NoiseThreshold} (which denote the fits obtained for $s=0\,$mJy using the envelope fitting method in Ref.~\cite{Paugnat2022}), including the fits obtained sporadically at $s>s_{\rm max}$.
That is, despite the jump in the inferred diameters $d_\varphi$ with $s=0\,$mJy (displayed in Fig.~\ref{fig:CirclipseInference}), the addition of substantial levels of instrument noise does not lead us to infer a different circlipse for any of the noise realizations.

Together, Figs.~\ref{fig:InferredDiameterDistributions} and~\ref{fig:NoiseThreshold} suggest that as the noise level increases, the RMSD also increases, so that the fits fail with higher frequency, as one might expect.
However, the resulting change in the distribution of the inferred diameters is, in relative terms, modest.
Thus, the main effect of increased instrument noise is a higher failure rate for measurements of the signal periodicity, but only a modest reduction in their accuracy and precision.

Above a certain noise threshold ($s_{\rm max}=3.2\,$mJy in our survey), the ability to reliably infer a periodicity from the signal becomes severely limited (i.e., for $s\geq s_{\rm max}$, the success rate for fitting attempts drops below $20\%$).
Finally, the presence of instrument noise \emph{alone} does not exacerbate transition-region effects nor lead one to infer a ``wrong'' circlipse.

In a real measurement, even if the noise level exceeds this threshold, it does not necessarily follow that the shape of the first photon ring is absent from the signal.
It may still be possible to extract the shape information using a more sophisticated method of data analysis that uses multiple observations and goes beyond the simple fitting method that is implemented herein.

\subsection{Suitable baseline windows}
\label{subsec:SuitableWindows}

To assess which baseline window is most suitable for a photon ring shape measurement, we now examine several profiles and fit their visibility amplitudes over various baseline windows, and then analyze the corresponding distribution of RMSDs.

More specifically, we perform fits using $270$ different time-averaged models, in which we vary the parameters of the underlying geometry (namely, the black hole spin and the observer inclination) and the three parameters $(\mu,\vartheta,\gamma)$ that characterize the radial emission profile \eqref{eq:JonhnsonSU}.
The precise values that we consider for these parameters are presented in Table~\ref{tbl:ProfileParameters}.
We systematically examine a set of baseline windows starting at $\cu{0,2,4,\ldots,100}\,\text{G}\lambda$ and extending over a range of $\cu{5,10,15,\ldots,50}\,\text{G}\lambda$.

\begin{table}
    \centering
    \begin{tabular}{cc}
    \hline 
    \hline 
    Parameter & Values \tabularnewline
    \hline 
    \hline 
    $a/M$ & $\cu{0\%,50\%,94\%}$ \tabularnewline
    $\theta_{\rm o}$ & $\cu{10^\circ,20^\circ,30^\circ}$ \tabularnewline
    $\mu$ & $\cu{r_-,\frac{1}{2}r_+,r_+,\frac{3}{2}r_+,2r_+} $\tabularnewline
    $\vartheta/M$ & $\cu{\frac{1}{2},1}$ \tabularnewline
    $\gamma$ & $\cu{-1,0,1}$ \tabularnewline
    \hline 
    \hline 
    \end{tabular}
    \caption{The $270$ profiles whose time-averaged images we survey in Fig.~\ref{fig:BaselineWindowSize} in search for the optimal baseline window for a shape measurement of the first photon ring.
    We consider three black hole spins, three observer inclinations, and $30$ emission profiles.
    This wide range of parameters accounts for many possible astrophysical scenarios in which the emission profile peaks at different locations and exhibits different levels of broadness and steepness.
    Although not all of these profiles are realistic, the parameter space is large enough to encompass a wide range of plausible astrophysical sources.
    The outer/inner event horizon radii are denoted by $r_\pm=M\pm\sqrt{M^2-a^2}$.}
    \label{tbl:ProfileParameters}
\end{table}

\begin{figure*}
	\centering
	\includegraphics[width=\textwidth]{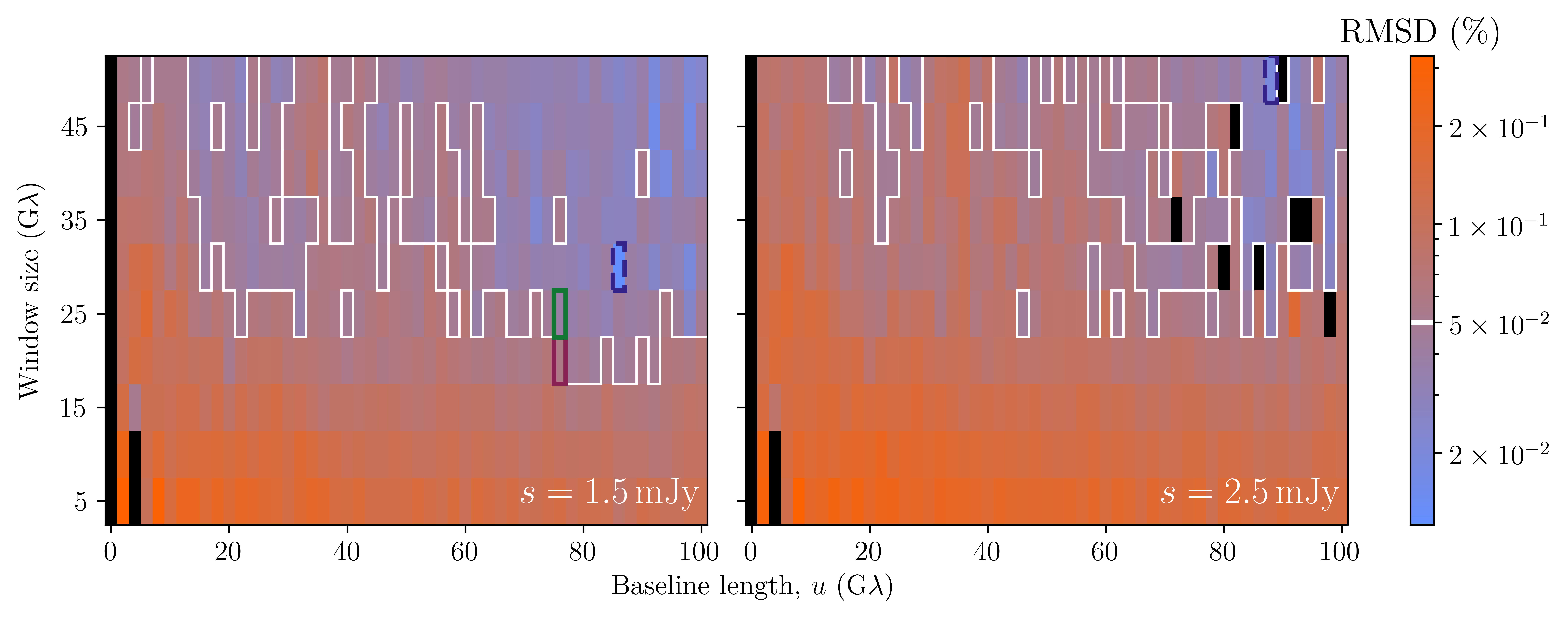}
	\caption{The distribution of the maximum value of the averaged RMSD obtained from the fits for the $270$ profiles given in Table~\ref{tbl:ProfileParameters}, in the presence of instrument noise levels $s=1.5\,$mJy (left) and $s=2.5\,$mJy (right), across baselines windows $[u_1,u_1+\Delta u]\,\text{G}\lambda$, where $u_1\in\cu{0,2,\ldots,100}$ and $\Delta u\in\cu{5,10,\ldots,50}$.
	The dashed purple cells denote the baseline windows $[86,116]\,\text{G}\lambda$ and $[88,138]\,\text{G}\lambda$, which have the lowest average worst-fit RMSD value for the cases $s=1.5\,$mJy and $s=2.5\,$mJy, respectively.
	The black cells correspond to baseline windows in which no fit is ever successful.
	The white contours correspond to a mean RMSD value of $0.05\%$, and they enclose $38\%$ and $19\%$ of the baseline windows in the cases $s=1.5\,$mJy and $s=2.5\,$mJy, respectively.}
	\label{fig:BaselineWindowSize}
\end{figure*}

This results in $51\times10=510$ baseline windows.
For each one of them, we perform $4$ fits per profile, each time adding a different realization of noise with $s=1.5\,$mJy or $s=2.5\,$mJy, producing a total of $1,101,600$ attempted fits.
For each profile across a given baseline window, we conservatively take the maximum RMSD of the $4$ fitting attempts.\footnote{When fitting in a given baseline window, sometimes a periodicity cannot be recovered from the visibility, so the fit fails and no RMSD is computed.
In some baseline windows (depicted by black cells in Fig.~\ref{fig:BaselineWindowSize}), \emph{all} of the fits fail.}
The result of this survey is presented in Fig.~\ref{fig:BaselineWindowSize}, where each baseline is represented by a cell whose color indicates the mean value of the $270$ RMSD maxima (a fairly conservative estimator).

The cells highlighted with dashed purple borders in Fig.~\ref{fig:BaselineWindowSize} correspond to the baseline windows $[86,116]\,\text{G}\lambda$ and $[88,138]\,\text{G}\lambda$, which have the lowest average RMSD when $s=1.5\,$mJy or $s=2.5\,$mJy, respectively.
The average RMSD of the fits across the window $[86,116]\,\text{G}\lambda$ is $0.012\%$, while the average RMSD across $[88,138]\,\text{G}\lambda$ is $0.019\%$.
The black cells indicate a complete inability to recover a fit: in the baseline windows corresponding to those cells, every fitting attempt is unsuccessful.
The white solid lines are contours corresponding to a mean RMSD value of $0.05\%$, and they enclose $38\%$ and $19\%$ of the baseline windows in the cases $s=1.5\,$mJy and $s=2.5\,$mJy, respectively.

To demonstrate the qualitative difference in the quality of the fits just above and below the (arbitrary) RMSD threshold of $0.05\%$, we present in App.~\ref{App:Threshold} examples of fits performed within the baseline windows highlighted with green and red borders in the left panel of Fig.~\ref{fig:BaselineWindowSize}.

We can see in Fig.~\ref{fig:BaselineWindowSize} that the more favorable baseline windows (i.e., those with a lower RMSD) are generally concentrated in the upper-right section of the heatmaps.
In other words, longer baselines and larger window sizes lead to better shape measurements.

The preference for larger window sizes is explained by the oscillatory behavior of the visibility amplitudes.
Each measurement requires finding the parameters that best fit an oscillatory function to some noisy data set.
Over a shorter window, and in the presence of noise, a larger proportion of the parameter space is consistent with the observed signal.
By contrast, observations over longer windows impose stronger constraints on the parameters of the fit, which must reproduce an increasingly specific oscillatory (and decaying) pattern over a larger domain.

The preference for longer baselines (within this region of the visibility domain and for the profiles considered) can be understood in two ways.
First, as the visibility amplitude decreases rapidly on baselines close to $0\,$mJy, fits on such baselines frequently fail (e.g., for both levels of noise, all the fits at baseline windows starting at $0\,\text{G}\lambda$ completely fail).
Second, as the baseline length increases, the visibility amplitude enters the universal photon ring regime and thus becomes increasingly independent of the astrophysical profile~\cite{JohnsonLupsasca2020,GLM2020}, resulting in sharper probes of the underlying geometry. 

However, when the baseline length increases too much, the steady decay of the power in the visibility amplitude leads to a decreased SNR, which brings down the quality of fits.
All else being equal, instrument noise generally has a stronger effect on larger baselines: by Eq.~\eqref{eq:UniversalVisibility}, the power decreases roughly like $\sqrt{u}$ with increasing baseline, so the SNR decreases like $\sqrt{u}$.
Since we only consider baselines extending up to $150\,\text{G}\lambda$ in this survey, this effect is not substantial, but if one were to consider even longer baselines, then a baseline-dependent noise level might be more appropriate than the fixed levels considered here.

Since SNR generally decreases with increasing baseline, it makes sense that regions of total failure appear in the ``better'' (longer) baseline windows in the $s=2.5\,$mJy case before appearing in the ``worse'' (shorter) windows.
As the noise level $s$ increases, we expect its effect to be strongest on those baselines where the SNR is smallest, that is, on longer baselines such as those appearing in the upper-right region of Fig.~\ref{fig:BaselineWindowSize}. 

The substantial increase in the fitting failure rate as the noise level grows from $s=1.5\,$mJy to $s=2.5\,$mJy is consistent with Fig.~\ref{fig:InferredDiameterDistributions}, and is a recurring feature in the surveys described in the rest of this paper.
Despite the significant failure rate at $s=2.5\,$mJy, there are still some baseline windows in the upper-right region of the plot whose corresponding RMSDs are comparable to the fits at noise level $s=1.5\,$mJy.
Indeed, the maximum RMSD for the fits across the window $[88,138]\,\text{G}\lambda$ with noise level $s=2.5\,$mJy is only $0.007\%$ larger than the maximum RMSD for the fits across the window $[86,116]\,\text{G}\lambda$ with noise level $s=1.5\,$mJy.

This is another recurring theme of our survey: in most baseline windows, as the instrument noise increases, the likelihood of detecting a periodic signal decreases, but, provided that it is detectable, its measurement accuracy does not vary much with the noise level.

As previously indicated, our chosen threshold RMSD of $0.05\%$ indicated by the white contours in Fig.~\ref{fig:BaselineWindowSize} is somewhat arbitrary (see App.~\ref{App:Threshold}).
We consider this choice to be rather conservative in the sense that it corresponds to qualitatively good circlipse fits for the photon ring shape. 
In a real space mission, the actual target RMSD value would depend on the science requirements.

If one can ignore orbital constraints and choose any baseline window to observe in, then Fig.~\ref{fig:BaselineWindowSize} indicates which windows are better suited to accurate measurements of the photon ring, \emph{on average}, for the $270$ profiles that we consider here.
This does not mean, for example, that in the presence of $s=1.5\,$mJy of instrument noise, all $270$ profiles considered are optimally observed across the window $[86,116]\,\text{G}\lambda$.
Indeed, for some profiles, a different baseline window might be optimal.
For the profile P1 (see Table~\ref{tbl:NominalCase}) considered throughout this paper, for example, the visibility amplitude in the window $[86,116]\,\text{G}\lambda$ lies in a transition region between regimes dominated by the $n=1$ and $n=2$ rings.
As such, a shorter baseline window may be preferable in order to avoid transition effects.
Besides, Fig.~\ref{fig:BaselineWindowSize} does not indicate the absolute quality of the fits across each window.
If a near-future experiment were to limit us to narrower baseline windows starting on shorter baselines, then according to Fig.~\ref{fig:NoiseThreshold}, accurate shape measurements of the photon ring would still be possible: the least favorable regions of Fig.~\ref{fig:BaselineWindowSize} have RMSDs comparable to those at the upper range shown in the left panel of Fig.~\ref{fig:NoiseThreshold}, and as the left panel shows, fits with these RMSDs have inferred diameters that remain close to those inferred in the absence of noise.

Thus far, we have only studied how instrument noise alone affects the extraction of a photon ring shape from the visibility amplitude of time-averaged images.
Next, we study the effects of astrophysical fluctuations on our ability to infer the photon ring shape. 

\section{Impact of astrophysical fluctuations}
\label{sec:AstrophysicalFluctuations}

We begin this section by considering how the scale of astrophysical fluctuations, which is characterized by the parameter $\sigma$ in Eq.~\eqref{eq:VariableSource}, can affect our ability to extract an interferometric diameter in visibility space.
We then consider the effect of the number $N$ of ``snapshots'' over which we time average the visibility.
This averaging can be carried out in one of two ways: either coherently, or incoherently.
Coherent averages retain phase information by first averaging the complex visibility and then taking the amplitude, whereas incoherent averages discard the relative phases by directly averaging at the level of the visibility amplitude.

As $N\rightarrow\infty$, astrophysical fluctuations wash out of the average over snapshot images, and thus one recovers the time-averaged image that corresponds to the radial emission profile \eqref{eq:JonhnsonSU} with the same $J_{\rm SU}(r_{\rm s})$ parameters.
A coherently averaged visibility corresponds to the visibility of this time-averaged image.
Experimentally, however, our current technology limits us to incoherent averaging.

\subsection{Impact of the number of snapshots averaged}

We consider two emissivity profiles, six magnitudes $\sigma$ of astrophysical fluctuations, and various spin-inclination combinations specified in Table~\ref{tbl:AstrophysicalFluctuations}, comprising a total of $108$ different profiles.
For each of these profiles, we ray trace a simulated black hole movie with a duration of $1000M$, sampled every $10M$ to produce a total of $N=100$ snapshots. We present the distribution of the power across the baseline window $[86, 116]\,\text{G}\lambda$ over all considered models in App.~\ref{App:Power}.

\begin{table}
    \centering
    \begin{tabular}{cc}
    \hline 
    \hline 
    Parameter & Values \tabularnewline
    \hline 
    \hline 
    $a/M$ & $\cu{2\%,50\%,94\%}$ \tabularnewline
    $\theta_{\rm o}$ & $\cu{10^\circ,20^\circ,30^\circ}$ \tabularnewline
    $\sigma$ & $\cu{0.0,0.05,0.15,0.25,0.35,0.45}$ \tabularnewline
    P$1=(\mu,\vartheta/M,\gamma)$ & $\pa{r_-,\frac{1}{2},-\frac{3}{2}}$ \tabularnewline
    P$2=(\mu,\vartheta/M,\gamma)$ & $\pa{\frac{3}{2}r_+,1,0}$ \tabularnewline
    \hline 
    \hline 
    \end{tabular}
    \caption{Parameters for the $108$ movies with astrophysical fluctuations.
    For each of the profiles in Table~\ref{tbl:NominalCase}, we consider three black hole spins, three observer inclinations, and six fluctuation scales $\sigma$.
    When $\sigma=0$, the source \eqref{eq:VariableSource} reduces to the time-averaged model \eqref{eq:JonhnsonSU} with the same parameters for the radial emission profile $J_{\rm SU}(r_{\rm s})$.
    The outer/inner event horizon radii are denoted by $r_\pm=M\pm\sqrt{M^2-a^2}$. }
    \label{tbl:AstrophysicalFluctuations}
\end{table}

For each of the $108$ profiles, and for $N=5$, $10$, $20$, and $70$ snapshots, we conduct $10$ fits, each time recording the maximum (i.e., worst-fit) value of the RMSD.
In each case, we randomly choose $N$ snapshots from the total of $100$, take either their coherent or incoherent average, and then attempt a fit to the resulting visibility amplitude.

\begin{figure*}
    \centering
    \includegraphics[width=0.95\columnwidth]{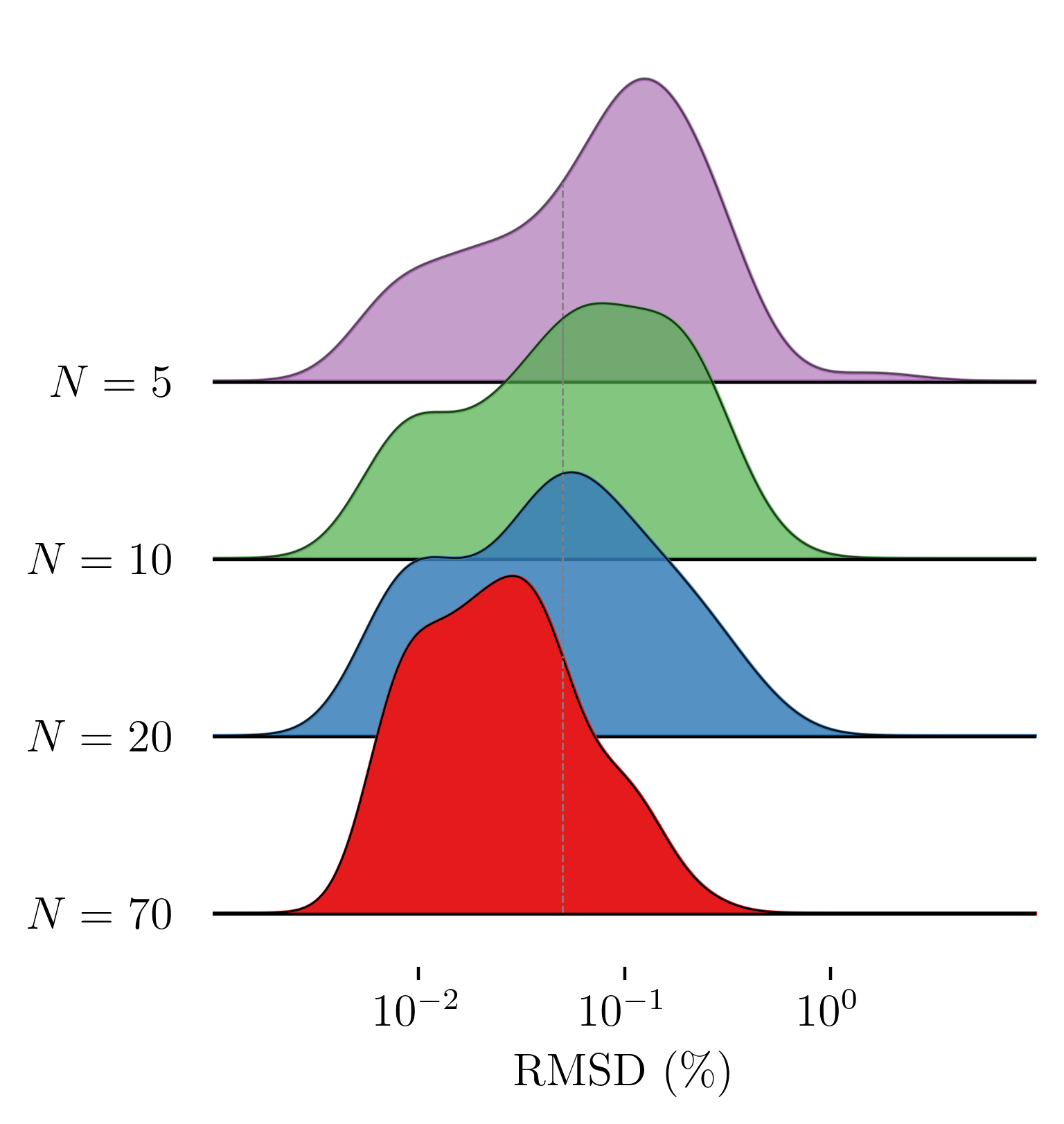}
\includegraphics[width=0.95\columnwidth]{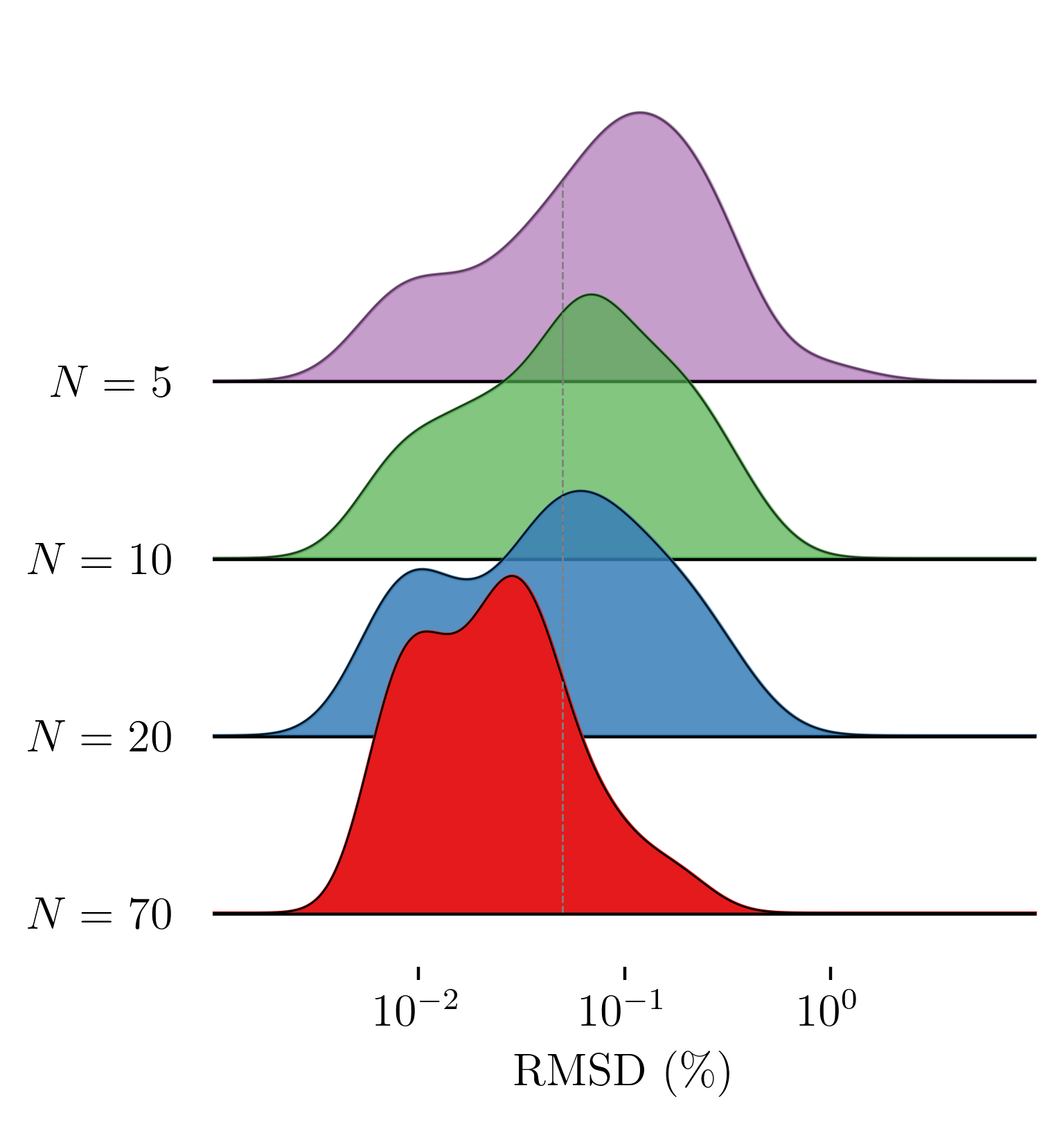}
    \caption{The distributions of the worst-fit RMSD among $10$ fits carried out for the $108$ time-dependent profiles given in Table~\ref{tbl:AstrophysicalFluctuations}, when taking the coherent (left), and incoherent (right) average of (from top to bottom) $N=5$ (purple), $10$ (green), $20$ (blue), and $70$ (red) snapshots.
    The vertical dashed line denotes an RMSD of $0.05\%$.
    As the number of snapshots $N$ increases, the RMSD of the fits generally decreases (i.e., the histograms shift toward the left) and the number of successful fits increases.}
    \label{fig:SnapshotAveraging}
\end{figure*}

The distributions of the worst-fit RMSD are shown in Fig.~\ref{fig:SnapshotAveraging} for the different values of $N$, for both coherent (left panel) and incoherent (right panel) averages.
Since these distributions are very similar, it follows that the phase variations among the images are relatively small, that SNR is high, and that the strong signal (large amplitude) features are dominating the snapshot images.

As expected, increasing the number $N$ of snapshots  that we average over causes the RMSD distributions in Fig.~\ref{fig:SnapshotAveraging} to shift toward lower values.
In particular, for the coherent averages, we find that $30\%$ of the profiles have RMSD values below our chosen threshold of $0.05\%$ (drawn as a vertical dashed line) when $N=5$.
At the other extreme, when $N=70$, this fraction grows to $59\%$.
As expected, we observe a similar trend for the incoherent averages.

Increasing the number $N$ of averaged snapshots not only lowers the worst-fit RMSD, but also improves our ability to detect a periodicity: for the coherent averages, the percentage of successful fits is $53\%$, $55\%$, $61\%$, and $66\%$ when the snapshots averaged over number $N=5$, $N=10$, $N=20$, and $N=70$, respectively, with similar values for the incoherent average.

These results are unsurprising and consistent with the fact that in the $N\to\infty$ limit, one recovers a time-averaged image dominated by a single persistent feature: the photon ring.
Increasing the number $N$ of snapshots lessens the impact of astrophysical fluctuations, reduces the RMSD of the fits, and enhances our ability to recover a periodicity.
However, the precise value of $N$ such that the RMSD dips below a certain threshold depends on the specific details of the fluctuations.

\subsection{Impact of the scale of astrophysical fluctuations}

Next, we survey the impact that the magnitude $\sigma$ of the astrophysical fluctuations has on our ability to infer an interferometric diameter.
For this survey, we keep all the other parameters fixed and vary only $\sigma$. 

\begin{figure*}
    \centering
    \includegraphics[width=\textwidth]{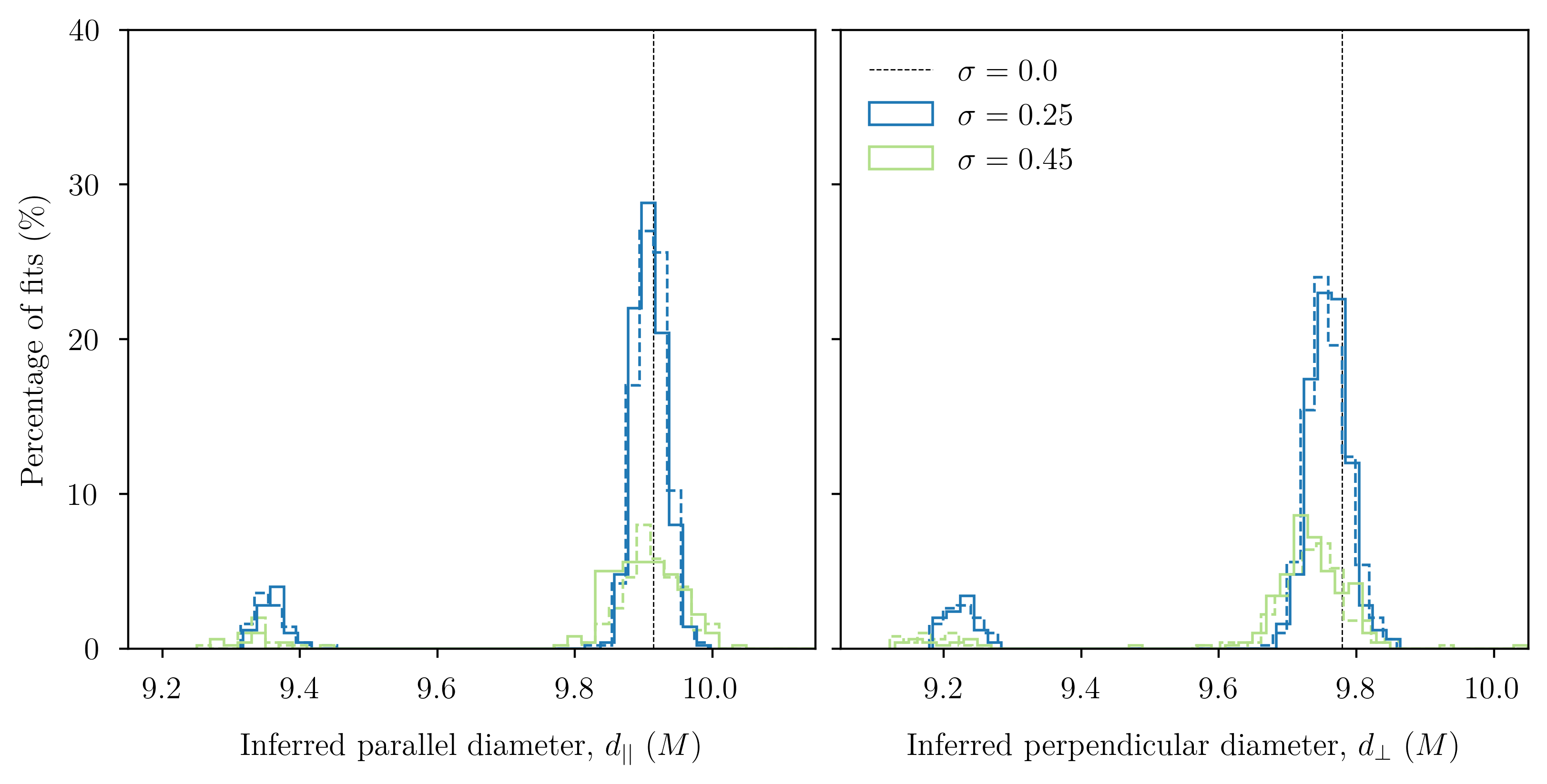}
    \caption{The multi-peaked distributions of diameters inferred from a time-dependent black hole movie with $\sigma=0.25$ (blue) or $\sigma=0.45$ (green) and underlying parameters given in Table~\ref{tbl:NominalCase}.
    We carry out $500$ fits across the baseline window $[86,116]\,\text{G}\lambda$ using a visibility amplitude produced by taking the coherent (solid lines) and incoherent (dashed lines) averages of $N=10$ snapshots randomly chosen in each fitting attempt.
    The dashed vertical lines represent the inferred diameters in the absence of astrophysical fluctuations ($\sigma=0$).
    As $\sigma$ increases, the centers and widths of the histograms do not undergo substantial changes, whereas their areas decrease significantly: in the coherently-averaged case, $95\%$ of the fits with $\sigma=0.25$ are successful, while that value decreases to $43\%$ when $\sigma=0.45$.
    For a small fraction of the randomly chosen sets of $N=10$ snapshots, we see the formation of a secondary peak corresponding to a circlipse with the same fractional asymmetry as the one closest to the shape of the critical curve, up to variations introduced by the instrument noise (see Fig.~\ref{fig:FractionalAsymmetry}).}
	\label{fig:AstrophysicalFluctuations}
\end{figure*}

We consider images of a black hole surrounded by the profile P1 given in Table~\ref{tbl:NominalCase}, and carry out $500$ fits across the baseline window $[86,116]\,\text{G}\lambda$, first with $\sigma=0.25$, and then again with $\sigma=0.45$.
Each fit is performed on a visibility amplitude obtained by averaging over $N=10$ snapshots randomly chosen during each fitting attempt.
We show the distributions of the inferred diameters $d_\parallel$ and $d_\perp$ resulting from coherent and incoherent averaging as the solid and dashed lines in Fig.~\ref{fig:AstrophysicalFluctuations}, respectively.
In these plots, the vertical dashed lines are the diameters inferred in the case $\sigma=0$ (i.e., with no astrophysical fluctuations), which are the same as those in Fig.~\ref{fig:InferredDiameterDistributions}. 

Unlike the distributions of diameters inferred in the presence of instrument noise (see Fig.~\ref{fig:InferredDiameterDistributions}), the diameters inferred in the presence of astrophysical fluctuations now form a multi-peaked distribution (as shown in Fig.~\ref{fig:AstrophysicalFluctuations}).
As discussed in the previous section, different realizations of instrument noise always lead to a measurement of the \emph{same} circlipse as in the case $s=0\,$mJy, up to a slight distortion and displacement.
By contrast, astrophysical fluctuations can change the structure of the image in ways that instrument noise cannot.
For instance, as the lower panel of Fig.~\ref{fig:AstrophysicalFluctuations} clearly shows, astrophysical fluctuation can introduce photon ring mimickers that can modify the periodicity in the visibility amplitude.
The impact of this effect on the inferred circlipse shape is exacerbated across windows located in regions of the visibility domain that correspond to transitions between rings.

The circlipse corresponding to the smaller peak in Fig.~\ref{fig:AstrophysicalFluctuations} is the one directly below the blue curve in the lower-left panel in Fig.~\ref{fig:MultiPeakedCirclipses}, which shows a typical fit with $\sigma=0.25$ in this survey.
This second peak arises because the fit is performed in a baseline window $[86, 116]\,\text{G}\lambda$, which for this particular profile lies within a transition region between the regimes dominated by the first and second photon rings.

We recall that in the $s=0\,$mJy and $\sigma=0$ case, the inferred diameters $d_\varphi$ jump between two circlipses (e.g., the circlipses starting at $9.2M$ and $9.8M$ in Fig.~\ref{fig:CirclipseInference}).
The upper-left panel of Fig.~\ref{fig:MultiPeakedCirclipses} shows that these circlipses have similar joint goodness-of-fits in the absence of noise and fluctuations: the joint goodness-of-fit of the ``best'' one (the blue curve in Fig.~\ref{fig:MultiPeakedCirclipses}) is $0.0023$, while that of the one directly below is $0.0016$.
These two values of the goodness-of-fit are ``close'' compared to the next highest value of $0.00046$ (corresponding to the circlipse staring at $10.3M$), a factor of $3.6$ times lower than the circlipse starting at $9.2M$.

Averaging over different sets of snapshots with non-zero astrophysical fluctuations leads to changes in the visibility amplitude to which we apply our fitting method.
In a small fraction of our snapshot averages, the visibility amplitude is such that this second circlipse has a larger goodness-of-fit than the one closest to the shape of the critical curve, which leads us to infer diameters that lie close to the smaller peaks in Fig.~\ref{fig:AstrophysicalFluctuations}.

\begin{figure}
    \centering
    \includegraphics[width=\columnwidth]{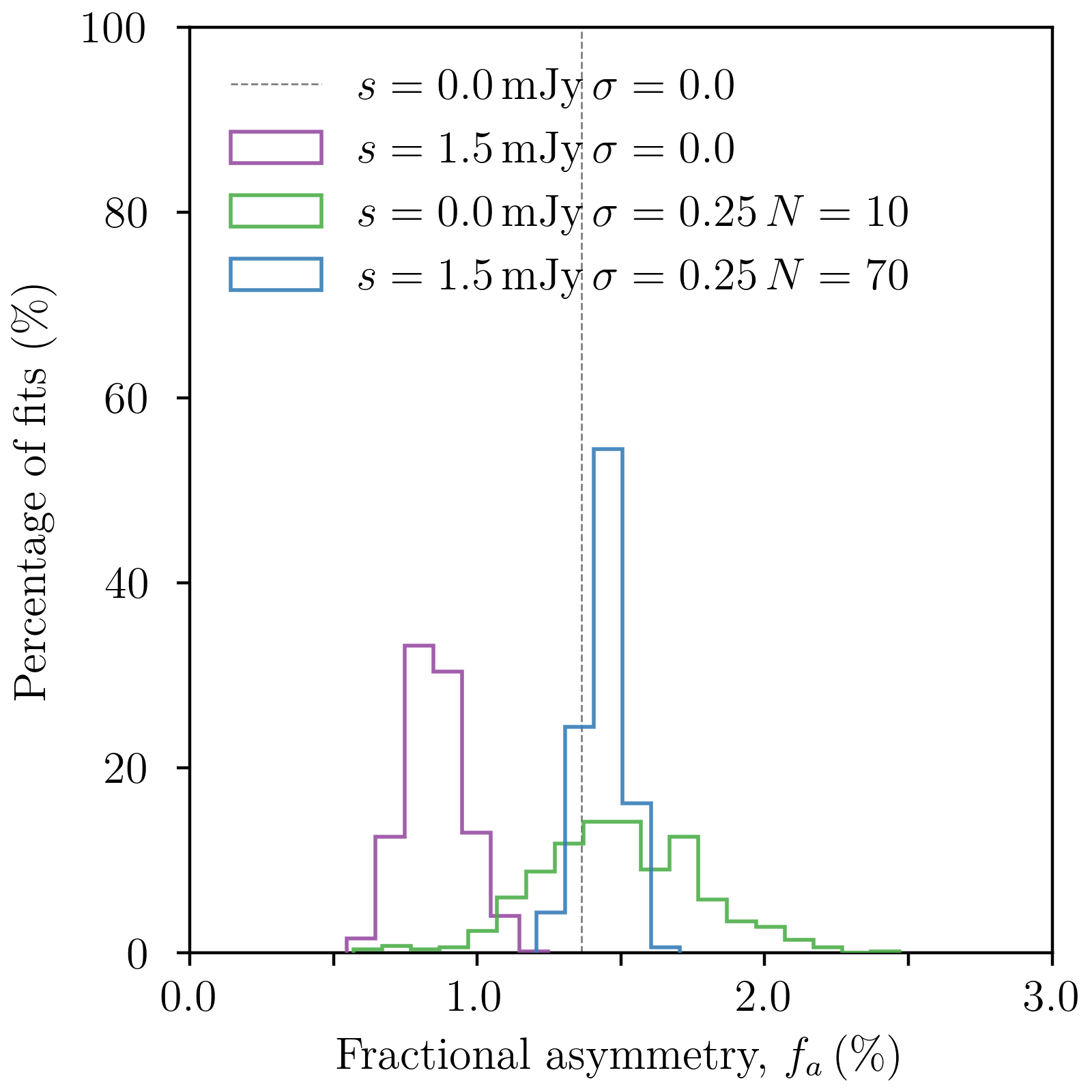}
    \caption{The distributions of the fractional asymmetry \eqref{eq:FractionalAsymmetry} inferred from simulations of a black hole surrounded with the emission profile P1 specified in Table~\ref{tbl:NominalCase}, and with various levels of instrument noise $s$ and astrophysical fluctuations $\sigma$.
    The vertical dashed line denotes the fractional asymmetry of the ``correct'' circlipse inferred in the absence of either noise or fluctuations (see Fig.~\ref{fig:CirclipseInference}).
    This figure includes the fractional asymmetry of all of the circlipses inferred in the presence of astrophysical fluctuations (see Figs.~\ref{fig:AstrophysicalFluctuations}  and~\ref{fig:RealisticMeasurement}), suggesting that even if one infers a different circlipse due to transition effects, nevertheless information about the fractional asymmetry may still be recovered from such an observation.}
\label{fig:FractionalAsymmetry}
\end{figure}

Interestingly, the green histogram in Fig~\ref{fig:FractionalAsymmetry} shows that the two circlipses inferred in this survey have the same fractional asymmetry, up to slight variations introduced by the fluctuations.

In other words, for a given geometry (i.e., for a fixed black hole spin and observer inclination), the distribution of the inferred fractional asymmetry is not multi-modal, while the inferred circlipses all differ from each other by some constant scaling factor.
As such, even though the diameters $d_\varphi$ of the secondary circlipses are farther away from the shape of the critical curve (whose overall size is set by the black hole mass), they nonetheless retain some information about the spacetime geometry, encoded in the ring asymmetry. 

Even when averaging over only $N=10$ snapshots, the distributions for the inferred diameters $d_\parallel$ and $d_\perp$ are quite closely aligned with the diameters inferred in the absence of astrophysical fluctuations, and indeed both distributions display a similar width.
That is, as $\sigma$ grows from zero to $\sigma=0.25$, and subsequently to $\sigma=0.45$, we do not observe a significant reduction in the accuracy or precision of the inferred diameters.
Rather, the most pronounced effect that we observe is a reduction in the area of the histograms, or equivalently, an increase of the failure rate for fitting attempts.
The percentages of successful fits for $\sigma=0.25$ and $\sigma=0.45$ are $95\%$ and $43\%$, respectively, with similar figures for the incoherent averages.

Consistent with the effects of instrument noise, we see that in the presence of increasingly large astrophysical fluctuations, our ability to detect a periodic signal in the visibility amplitude is significantly impaired, while the accuracy and precision of our fits, when we are able to obtain them, does not decrease nearly as substantially. 

When running these simulations, we first specify the geometrical parameters $(a,\theta_{\rm o})$, from which we compute the critical curve analytically.
Thus, we know \emph{a priori} that the dominant peak in Fig.~\ref{fig:AstrophysicalFluctuations} is that of the ``correct'' circlipse, that is, the one closest to the shape of the (in itself unobservable) critical curve (see, e.g., Fig.~\ref{fig:CirclipseInference}).

The distance between the two inferred circlipses is approximately equal to $1/u_w$, where $u_w$ represents the length of the baseline in the middle of the window \cite{Paugnat2022}.
Thus, in this scenario, the size difference between the two circlipses is only a few percent.
Therefore, in a realistic observation, one could perhaps identify the ``correct'' peak provided one had a strong prior on the mass-to-distance ratio, as is the case for Sgr~A*.

\section{Astrophysical fluctuations and instrument noise}
\label{sec:NoiseAndFluctuations}

Having considered the effects of instrument noise and astrophysical fluctuations separately, we now study their combined effect on our ability to infer an interferometric diameter.
For each of the astrophysical profiles listed in Table~\ref{tbl:NominalCase}, we attempt $500$ fits across the baseline window $[86,116]\,\text{G}\lambda$ after averaging over $N=5$, $10$, $20$, and $70$ snapshots generated from a movie with fluctuations of magnitude $\sigma=0.25$.

\begin{figure*}
    \centering
    \includegraphics[width=\textwidth]{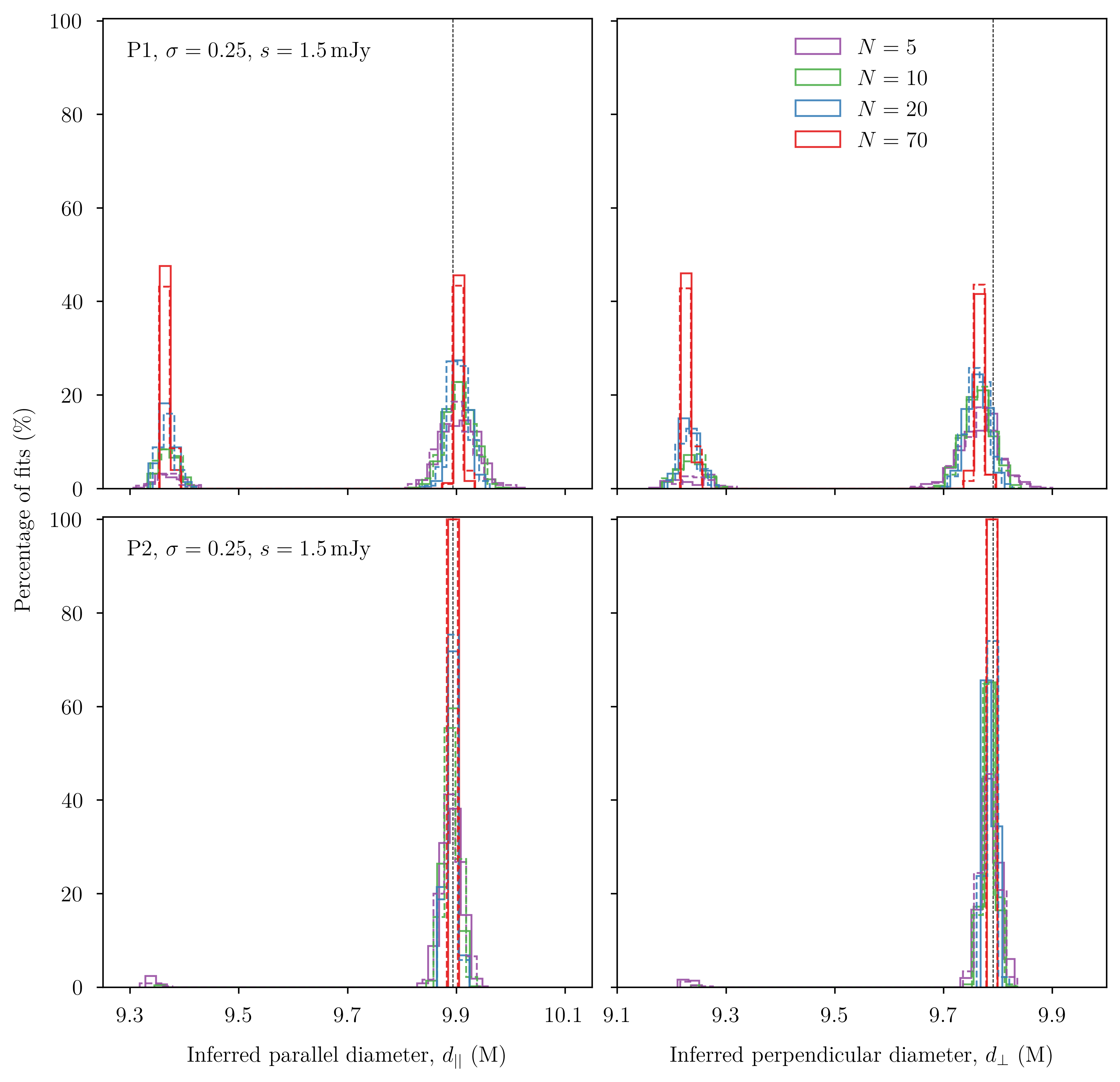}
    \caption{The distributions of diameters inferred from a time-dependent movie of a black hole surrounded with the emission profile P1 (upper panel) or P2 (lower panel) and underlying parameters given in Table~\ref{tbl:NominalCase}, in the presence of a level $s=1.5\,$mJy of instrument noise and $\sigma=0.25$ of astrophysical fluctuations.
    We carry out $500$ fits across the baseline window $[86,116]\,\text{G}\lambda$ using a visibility amplitude produced by taking the coherent (solid lines) and incoherent (dashed lines) averages of $N=5$ (purple), $N=10$ (green), $N=20$ (blue), and $N=70$ (red) snapshots randomly chosen in each fitting attempt.
    The dashed vertical lines represent the diameters inferred in the absence of instrument noise and astrophysical fluctuations (i.e., $s=0\,$mJy and $\sigma=0$).
    For the profile P1, the baseline window lies (at every angle $\varphi$) within a transition region between the regimes dominated by the first and second photon rings.
    As a result, the distributions are multi-peaked and, for $N=70$, of roughly the same height.
    For the profile P2, which lies in the regime dominated by the $n=1$ ring for most angles $\varphi$, the distributions effectively have a single peak that corresponds to the interferometric diameter $d_\varphi^{(1)}$ of the ring.
    For both profiles, as the number $N$ of snapshots increases, the histograms become narrower and the percentage of successful fits increases significantly.}
\label{fig:RealisticMeasurement}
\end{figure*}

In each fit, we add a separate realization of instrument noise with $s=1.5\,$mJy to each of the $N$ randomly chosen snapshots, and then take their coherent and incoherent average to obtain a time-averaged visibility amplitude to which we apply our fitting method.
We present the resulting inferred-diameter distributions for the coherent (solid histograms) and incoherent (dashed histograms) averages in Fig.~\ref{fig:RealisticMeasurement}.
The dashed vertical lines therein represent the inferred diameters in the case of $s=0\,$mJy and $\sigma=0$, and are identical to those in Figs.~\ref{fig:InferredDiameterDistributions} and~\ref{fig:AstrophysicalFluctuations}.

For the profile P1, the inferred-diameter distributions in Fig.~\ref{fig:RealisticMeasurement} are multi-peaked, with two peaks of similar heights separated by $\Delta d_\phi\approx 1/u_w$, where $u_w$ is the length of the baseline in the middle of the sampled window.
This suggests that in the presence of fluctuations, the addition of instrument noise exacerbates any transition effects and leads to the inference of a wrong circlipse almost equally as it does the correct one.
By contrast, for the profile P2, the visibility amplitude is not always sampled in a transition region at each of the $36$ baseline angles (see, e.g., the upper panel of Fig.~4 in Ref.~\cite{CardenasAvendano2023}).
As a result, the inferred-diameter distributions for profile P2 have effectively only one peak.
Indeed, Fig.~\ref{fig:RealisticMeasurement} shows that already with $N=20$ snapshots, one always infers the same, correct circlipse.

This suggests that in an actual observation, if one were to obtain two such peaks by randomly sampling different subsets of the measured data, then one might be led to conclude that the visibility amplitude was measured in a transition region.
But even if this happens, and one does not have a prior that is sufficiently strong to break the degeneracy, Fig.~\ref{fig:FractionalAsymmetry} suggests that the measured diameter $d_\varphi$ still retains correct information about the fractional asymmetry of the photon ring.

For the profile P1, Fig.~\ref{fig:RealisticMeasurement} shows that averaging has a very strong effect, even with only $N=5$ snapshots: the peaks of the (purple) $N=5$ histograms corresponding to the correctly inferred circlipse are more closely aligned with the dashed vertical lines than are the $s=1.5\,$mJy histograms in Fig.~\ref{fig:InferredDiameterDistributions}, even if astrophysical fluctuations are now present.
For $\sigma=0.45$, however, the distributions of diameters inferred from $N=5$ snapshots are quite wide, indicating that more snapshots are required to tame this level of fluctuation.

As the scale of astrophysical fluctuations and the level of instrument noise grow larger, the dominant effect for profile P1 is a reduction in the areas of the histograms.
The $N=5$ histograms for $\sigma=0.25$ have much smaller areas than the $s=1.5\,$mJy histograms in Fig.~\ref{fig:InferredDiameterDistributions}.
More precisely, in the presence of only instrument noise with a level $s=1.5\,$mJy, $95\%$ of the fits are successful, but this figure decreases to $76\%$ (or $26\%$) after the addition of astrophysical fluctuations with $\sigma=0.25$ (or $\sigma=0.45$) and after averaging over only $N=5$ snapshots.
These figures, however, increase substantially as $N$ increases: for $N=70$ snapshots, $100\%$ of the fits with $\sigma=0.25$ are successful, and likewise, out of the $500$ fitting attempts for $\sigma=0.45$, only one fails.
However, for the profile P2, even with only $N=5$ snapshots, $99\%$ (or $83\%$) of the fits are successful in the presence of fluctuations at a scale $\sigma=0.25$ (or $\sigma=0.45$): as expected, it is easier to detect a periodicity outside of a transition region, and this is in fact possible even with as few as $N=5$ snapshots.

In summary, the combination of instrument noise and astrophysical fluctuations leads to two main effects: they make it harder to detect a periodicity in the visibility amplitude, and they exacerbate transition effects.
When observing in a transition region, the inferred-diameter distributions may become double-peaked.
This results in an ambiguity in the overall scale of the angle-dependent ring diameter $d_\varphi$, but fortuitously, not in its fractional asymmetry.
Outside of a transition region, the inferred-diameter distributions effectively have only one peak.
In either case, averaging over snapshots is a powerful tool: as $N$ increases, the periodicity of the average visibility amplitude becomes increasingly pronounced, even when both astrophysical fluctuations and instrument noise are present, and the inferred-diameter distributions become increasingly narrow.
In other words, our survey shows that, using only our simple data analysis method, it is possible to measure the \textit{shape} of the photon ring quite accurately and precisely, even though transitions effects may result in a loss of information about the overall ring scale, and thus make parameter inference more difficult.

\section{Conclusions}
\label{sec:Conclusion}

We have conducted a systematic study to investigate how instrument noise and astrophysical fluctuations can impact our ability to infer the shape of the photon ring.
Our findings reveal that it is feasible to detect a periodic signal in the visibility amplitude up to a certain threshold level of noise and astrophysical fluctuations.

Remarkably, up to this point, neither the instrument noise nor the fluctuations significantly compromise the accuracy or precision of our simulated measurements.
However, our analysis also identifies a critical threshold---a sharp transition---beyond which such a measurement becomes intractable with our fitting method.

Measuring an interferometric diameter and comparing it to the GR prediction is challenging in the presence of both instrument noise and source fluctuations.
However, time-averaging over several snapshots is a potent means for restoring the feasibility of a photon ring measurement.
Although our analysis underscores the formidable impact of instrument noise and astrophysical fluctuations, we do not rule out the possibility that a more sophisticated data analysis scheme could still detect a periodic signal in the visibility amplitude, even when our method could not.
Thus, further studies of data analysis techniques may present avenues for overcoming the limitations imposed by instrument noise and help refine our understanding of future observations.

It is especially challenging to infer an interferometric diameter from observations of the visibility amplitude in a baseline window where the signal of the first photon ring does not always dominate.
For example, sampling the visibility in a transition region between two regimes dominated by different rings can sometimes result in the inference of another circlipse.
Such a secondary circlipse may differ from the one whose shape is closest to that of the critical curve by an overall scaling factor, but nonetheless retains the same fractional asymmetry (up to noise effects).
In other words, even when observations are performed in a transition region, it remains possible to measure a photon ring shape with the correct fractional asymmetry, even if the inferred diameter is in some cases off by a constant.
Such effects can be mitigated using various consistency checks.
For instance, when analyzing a sequence of images of a black hole, one would expect consistent results across different sets of snapshots, which should lead to the inference of the same circlipse shape or consistent diameters at every baseline angle.
However, should discrepancies arise, they could serve as potential indicators that the observations have been performed in a transition region, and hence that their analysis requires a more sophisticated approach.

This first study has only scratched the surface of the difficulties inherent in measurements of the photon ring shape amidst noise and fluctuations.
This research is still in its early stages, and there are many ways to improve and refine our approach.

First, we have relied on a particular statistical method for modeling astrophysical fluctuations.
This approach has allowed us to simulate various emission models in a less computationally expensive way compared to running GRMHD simulations.
The fluctuations in our models all exhibit a correlation structure governed by the highly versatile Mat\'ern covariance function, which enjoys broad applicability and can even describe free scalar fields in Euclidean field theory~\cite{CardenasAvendano2022}.

By assessing whether the Mat\'ern covariance function accurately reproduces the correlation structure observed in GRMHD simulations, we not only probe the intricacies of turbulence in accretion flows but also strive to bridge the gap between mathematical models and real physical systems.
Varying this covariance is an exciting prospect for future exploration.

Our rather simplistic approach for the incorporation of instrument noise in simulated observations---consisting of the addition of Gaussian complex noise---paves the way for future analyses that will consider additional factors.

Modeling detailed features of future VLBI observations with a radio dish in space (such as its size and orbit), along with real-world conditions (such as weather) and a realistic array of multiple ground-based stations promises to refine our understanding of observational constraints.
Some of these aspects have been recently studied using GRMHD simulations and several possible configurations for an orbiting satellite in Ref.~\cite{Shlentsova2024}, which also found that space-VLBI missions can be used to image the variability in a black hole environment and its extended jets.

Currently planned space missions will target M87$^*$ and the supermassive black hole at the center of our galaxy, Sgr~A*.
Observations of Sgr~A* present more challenges posed by the propagation of the emitted light through the interstellar medium, and by the shorter timescale for variability in the source.
In particular, phenomena such as scattering cannot be neglected, which necessitates a reassessment of our methodologies and assumptions.

This work is a proof of concept for critical components in the data analysis pipeline that are essential to future missions targeting black hole photon rings.
Our results are part of the ongoing effort to transform black hole images into precision probes of fundamental phenomena in astrophysics, and offer encouraging prospects for future measurements of black hole parameters via photon ring observations.

\acknowledgments

We thank George Wong for many helpful discussions.
We are grateful to William and Kacie Snellings for their generous support.
A.C.-A. also acknowledges support from the Simons Foundation, and A.L. is supported in part by NSF grants AST-2307888 and PHY-2340457.
The simulations presented in this article were performed on computational resources managed and supported by Princeton Research Computing, a consortium of groups that includes the Princeton Institute for Computational Science and Engineering (PICSciE), and the Office of Information Technology's High Performance Computing Center and Visualization Laboratory at Princeton University.

\appendix

\section{Fits across the RMSD threshold}
\label{App:Threshold}

In Sec.~\ref{subsec:SuitableWindows}, we chose a threshold RMSD of $0.05\%$ to draw as a white contour in Fig.~\ref{fig:BaselineWindowSize}.
In our survey, we find that an RMSD of $0.05\%$ provides a conservative value for our simulated observations.
To illustrate this, we pick a particular profile (black hole spin $a/M=94\%$, observer inclination $\theta_{\rm o}=10^\circ$, and emission profile parameters $\mu=r_+$, $\vartheta=M$, and $\gamma=0$) and perform fits across two arbitrarily chosen baseline windows, $[76,96]\,\text{G}\lambda$ and $[76,101]\,\text{G}\lambda$, which are enclosed by the orange and red borders in Fig.~\ref{fig:BaselineWindowSize}, respectively, and lie on either side of the threshold RMSD region.

\begin{figure}
    \includegraphics[width=\columnwidth]{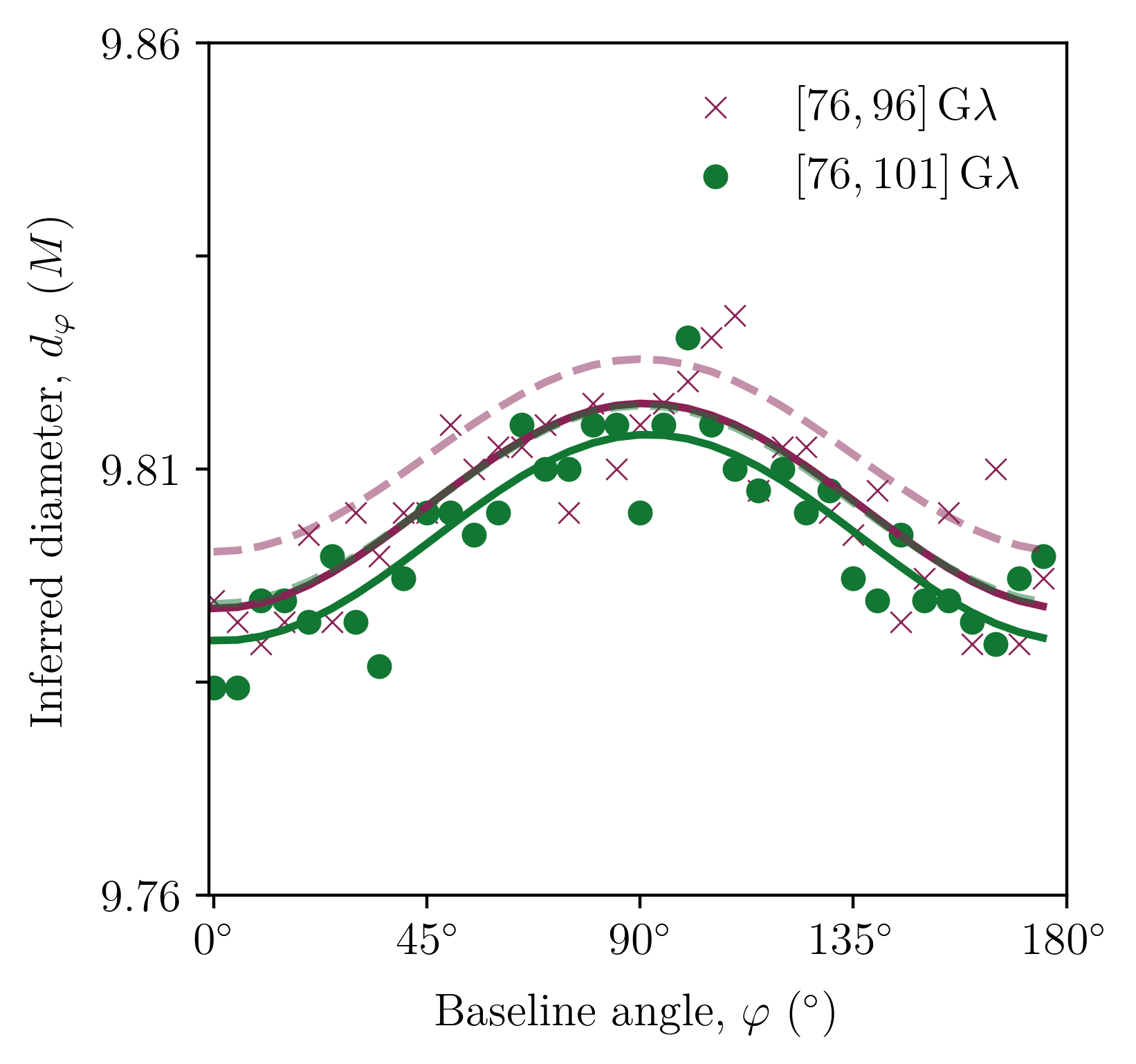}
    \caption{Inferred photon ring shapes \eqref{eq:Circlipse} for a black hole with spin $a/M=94\%$ surrounded by a radial emission profile \eqref{eq:JonhnsonSU} with parameters $\mu=r_+$, $\vartheta=M$, and $\gamma=0$, and observed from an inclination $\theta_{\rm o}=10^\circ$, in the presence of instrument noise levels $s=0\,$mJy (dashed lines) and $s=1.5\,$mJy (solid lines), across the baseline windows $[76,96]\,\text{G}\lambda$ (purple crosses) and $[76,101]\,\text{G}\lambda$ (green solid points).
    On these baselines, the visibility amplitude attains a minimum power \eqref{eq:Power} of $1.8\,$mJy and $2.3\,$mJy, and an average power of $2.0\,$mJy and $2.4\,$mJy, respectively.
    The RMSDs for these fits are $0.058\%$ (purple solid lines) and $0.048\%$ (green solid line), so they lie just above and below the critical RMSD of $0.05\%$, respectively.}
    \label{fig:CirclipseFits}
\end{figure}

The minimum and average power \eqref{eq:Power} across the first window are $1.8\,$mJy and $2.3\,$mJy, respectively, and $2.0\,$mJy and $2.4\,$mJy across the second window.
In Fig.~\ref{fig:CirclipseFits}, we show the best-fit circlipses obtained from a single fit across these baseline windows in the absence of noise, i.e., $s=0\,$mJy (dashed lines), and then in the presence of instrument noise at a level $s=1.5\,$mJy (solid lines).
The RMSD of the fits across $[76,96]\,\text{G}\lambda$ and $[76,101]\,\text{G}\lambda$ are $0.058\%$ and $0.048\%$, respectively.
The fit across $[76,96]\,\text{G}\lambda$ is markedly worse than the one across $[76,101]\,\text{G}\lambda$, with the inferred diameters for the former deviating more from its circlipse than the latter.
Indeed, the best-fit circlipses inferred in the presence of noise (solid lines) exhibit more significant deviations from their corresponding noiseless analogues for the fit across $[76,96]\,\text{G}\lambda$ (dashed purple line) than for the fit across $[76,101]\,\text{G}\lambda$ (green solid line).

As an indication of the quality of fit for this profile, for the fit across $[76,101]\,\text{G}\lambda$, the percentage change in the inferred diameters $d_\perp$ and $d_\parallel$ when going from $s=0\,$mJy to $s=1.5\,$mJy is $0.044\%$ and $0.035\%$, respectively, while the inferred diameters of the first ring differ from those of the critical curve by $0.40\%$ and $0.28\%$, respectively.

\section{Power in the baseline window \texorpdfstring{$[86,116]\,\text{G}\lambda$}{[86,116] Gλ} across all models}
\label{App:Power}

\begin{figure}
    \centering
\includegraphics[width=\columnwidth]{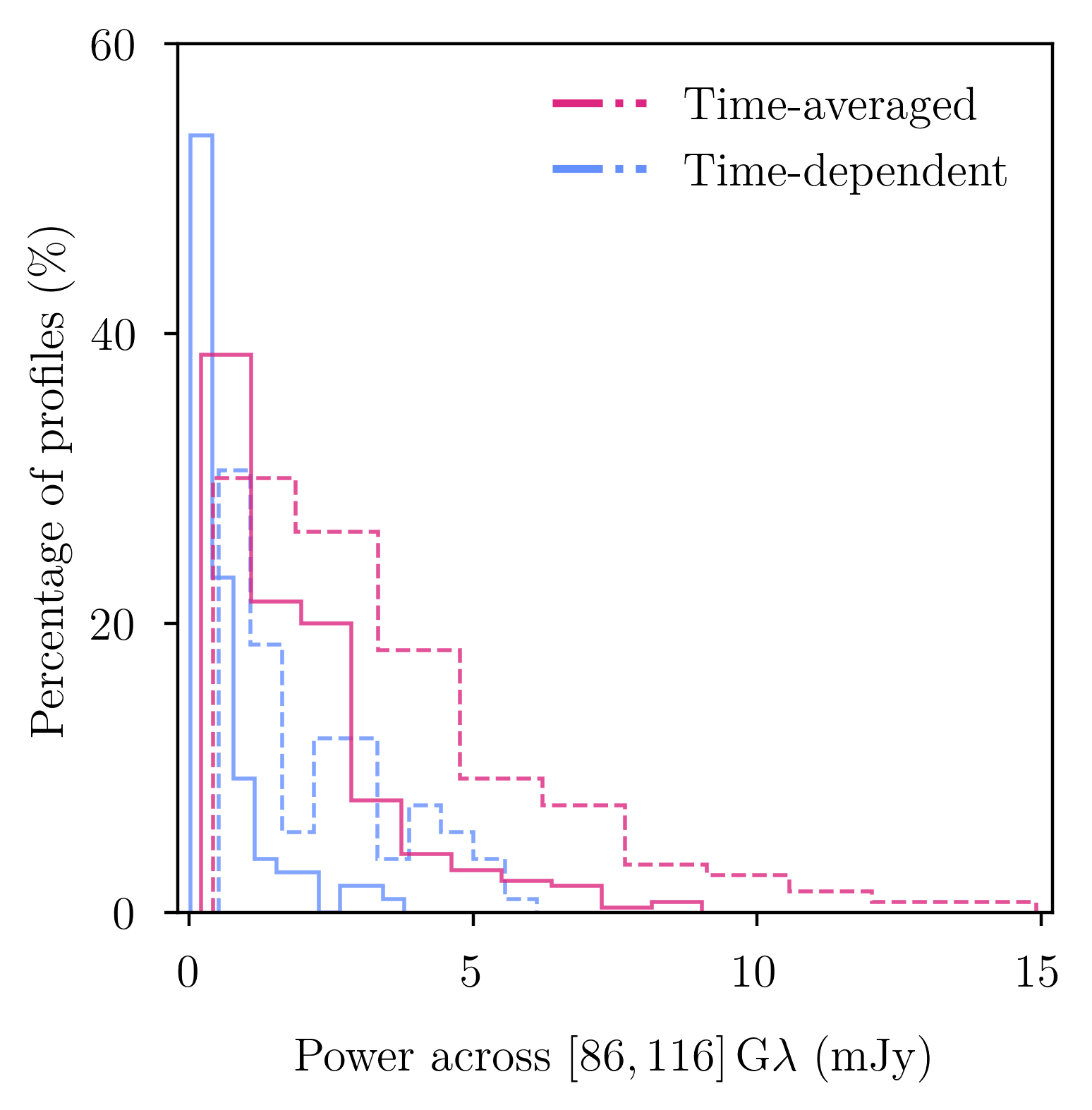}
    \caption{Distribution of the minimum (solid histograms) and average (dashed histograms) power \eqref{eq:Power} across the baseline window $[86,116]\,\text{G}\lambda$ for the $270$ time-averaged profiles (pink) given in Table~\ref{tbl:ProfileParameters}, and for the $108$ time-dependent profiles with astrophysical fluctuations (light blue) given in Table~\ref{tbl:AstrophysicalFluctuations}.
    For the time-averaged profiles, the minimum power attained at any baseline angle is $0.2\,$mJy, while the average power across all the baselines windows and all the profiles is $3.7\,$mJy.
    For the time-dependent profiles, the minimum and average powers are $0.03\,$mJy and $2.2\,$mJy, respectively.}
    \label{fig:PowerDistribution}
\end{figure}

To characterize the strength of the visibility amplitude across some given baseline window and at some baseline angle $\varphi$, we show in Fig.~\ref{fig:PowerDistribution} the distribution of the power \eqref{eq:Power} across the baseline window $[86,116]\,\text{G}\lambda$ for all the profiles listed in Tables~\ref{tbl:ProfileParameters} and~\ref{tbl:AstrophysicalFluctuations}.
The notion of power is especially important when thinking about the physical effects of some noise level, since, in combination with $s$, it provides a rough estimate of the SNR. 

From Fig.~\ref{fig:PowerDistribution}, one can see that the instrument noise levels $s=1.5\,$mJy and $s=2.5\,$mJy considered in the main text should indeed have a significant effect on the underlying visibility signal (as the right panels of Fig.~\ref{fig:MultiPeakedCirclipses} also show).

\bibliographystyle{apsrev4-1}
\bibliography{CKL}

\begin{thebibliography}{32}%
\makeatletter
\providecommand \@ifxundefined [1]{%
 \@ifx{#1\undefined}
}%
\providecommand \@ifnum [1]{%
 \ifnum #1\expandafter \@firstoftwo
 \else \expandafter \@secondoftwo
 \fi
}%
\providecommand \@ifx [1]{%
 \ifx #1\expandafter \@firstoftwo
 \else \expandafter \@secondoftwo
 \fi
}%
\providecommand \natexlab [1]{#1}%
\providecommand \enquote  [1]{``#1''}%
\providecommand \bibnamefont  [1]{#1}%
\providecommand \bibfnamefont [1]{#1}%
\providecommand \citenamefont [1]{#1}%
\providecommand \href@noop [0]{\@secondoftwo}%
\providecommand \href [0]{\begingroup \@sanitize@url \@href}%
\providecommand \@href[1]{\@@startlink{#1}\@@href}%
\providecommand \@@href[1]{\endgroup#1\@@endlink}%
\providecommand \@sanitize@url [0]{\catcode `\\12\catcode `\$12\catcode
  `\&12\catcode `\#12\catcode `\^12\catcode `\_12\catcode `\%12\relax}%
\providecommand \@@startlink[1]{}%
\providecommand \@@endlink[0]{}%
\providecommand \url  [0]{\begingroup\@sanitize@url \@url }%
\providecommand \@url [1]{\endgroup\@href {#1}{\urlprefix }}%
\providecommand \urlprefix  [0]{URL }%
\providecommand \Eprint [0]{\href }%
\providecommand \doibase [0]{http://dx.doi.org/}%
\providecommand \selectlanguage [0]{\@gobble}%
\providecommand \bibinfo  [0]{\@secondoftwo}%
\providecommand \bibfield  [0]{\@secondoftwo}%
\providecommand \translation [1]{[#1]}%
\providecommand \BibitemOpen [0]{}%
\providecommand \bibitemStop [0]{}%
\providecommand \bibitemNoStop [0]{.\EOS\space}%
\providecommand \EOS [0]{\spacefactor3000\relax}%
\providecommand \BibitemShut  [1]{\csname bibitem#1\endcsname}%
\let\auto@bib@innerbib\@empty
\bibitem [{\citenamefont {{Verbiest}}\ \emph {et~al.}(2021)\citenamefont
  {{Verbiest}}, \citenamefont {{Os{\l}owski}},\ and\ \citenamefont
  {{Burke-Spolaor}}}]{Verbiest2021}%
  \BibitemOpen
  \bibfield  {author} {\bibinfo {author} {\bibfnamefont {J.~P.~W.}\
  \bibnamefont {{Verbiest}}}, \bibinfo {author} {\bibfnamefont
  {S.}~\bibnamefont {{Os{\l}owski}}}, \ and\ \bibinfo {author} {\bibfnamefont
  {S.}~\bibnamefont {{Burke-Spolaor}}},\ }in\ \href {\doibase
  10.1007/978-981-15-4702-7_4-1} {\emph {\bibinfo {booktitle} {Handbook of
  Gravitational Wave Astronomy}}}\ (\bibinfo  {publisher} {Springer},\ \bibinfo
  {year} {2021})\ p.~\bibinfo {pages} {4}\BibitemShut {NoStop}%
\bibitem [{\citenamefont {{Agazie}}\ \emph {et~al.}(2023)\citenamefont
  {{Agazie}} \emph {et~al.}}]{NANOGrav2023}%
  \BibitemOpen
  \bibfield  {author} {\bibinfo {author} {\bibfnamefont {G.}~\bibnamefont
  {{Agazie}}} \emph {et~al.} (\bibinfo {collaboration} {NANOGrav
  Collaboration}),\ }\href {\doibase 10.3847/2041-8213/acdac6} {\bibfield
  {journal} {\bibinfo  {journal} {\apjl}\ }\textbf {\bibinfo {volume} {951}},\
  \bibinfo {eid} {L8} (\bibinfo {year} {2023})},\ \Eprint
  {http://arxiv.org/abs/2306.16213} {arXiv:2306.16213 [astro-ph.HE]}
  \BibitemShut {NoStop}%
\bibitem [{\citenamefont {{Johnson}}\ \emph {et~al.}(2020)\citenamefont
  {{Johnson}}, \citenamefont {{Lupsasca}} \emph
  {et~al.}}]{JohnsonLupsasca2020}%
  \BibitemOpen
  \bibfield  {author} {\bibinfo {author} {\bibfnamefont {M.~D.}\ \bibnamefont
  {{Johnson}}}, \bibinfo {author} {\bibfnamefont {A.}~\bibnamefont
  {{Lupsasca}}},  \emph {et~al.},\ }\href {\doibase 10.1126/sciadv.aaz1310}
  {\bibfield  {journal} {\bibinfo  {journal} {Science Advances}\ }\textbf
  {\bibinfo {volume} {6}},\ \bibinfo {pages} {eaaz1310} (\bibinfo {year}
  {2020})},\ \Eprint {http://arxiv.org/abs/1907.04329} {arXiv:1907.04329
  [astro-ph.IM]} \BibitemShut {NoStop}%
\bibitem [{\citenamefont {{Gralla}}\ \emph {et~al.}(2020)\citenamefont
  {{Gralla}}, \citenamefont {{Lupsasca}},\ and\ \citenamefont
  {{Marrone}}}]{GLM2020}%
  \BibitemOpen
  \bibfield  {author} {\bibinfo {author} {\bibfnamefont {S.~E.}\ \bibnamefont
  {{Gralla}}}, \bibinfo {author} {\bibfnamefont {A.}~\bibnamefont
  {{Lupsasca}}}, \ and\ \bibinfo {author} {\bibfnamefont {D.~P.}\ \bibnamefont
  {{Marrone}}},\ }\href {\doibase 10.1103/PhysRevD.102.124004} {\bibfield
  {journal} {\bibinfo  {journal} {\prd}\ }\textbf {\bibinfo {volume} {102}},\
  \bibinfo {eid} {124004} (\bibinfo {year} {2020})},\ \Eprint
  {http://arxiv.org/abs/2008.03879} {arXiv:2008.03879 [gr-qc]} \BibitemShut
  {NoStop}%
\bibitem [{\citenamefont {{Gralla}}\ \emph {et~al.}(2019)\citenamefont
  {{Gralla}}, \citenamefont {{Holz}},\ and\ \citenamefont
  {{Wald}}}]{Gralla2019}%
  \BibitemOpen
  \bibfield  {author} {\bibinfo {author} {\bibfnamefont {S.~E.}\ \bibnamefont
  {{Gralla}}}, \bibinfo {author} {\bibfnamefont {D.~E.}\ \bibnamefont
  {{Holz}}}, \ and\ \bibinfo {author} {\bibfnamefont {R.~M.}\ \bibnamefont
  {{Wald}}},\ }\href {\doibase 10.1103/PhysRevD.100.024018} {\bibfield
  {journal} {\bibinfo  {journal} {\prd}\ }\textbf {\bibinfo {volume} {100}},\
  \bibinfo {eid} {024018} (\bibinfo {year} {2019})},\ \Eprint
  {http://arxiv.org/abs/1906.00873} {arXiv:1906.00873 [astro-ph.HE]}
  \BibitemShut {NoStop}%
\bibitem [{\citenamefont {{Gralla}}\ and\ \citenamefont
  {{Lupsasca}}(2020{\natexlab{a}})}]{GrallaLupsasca2020a}%
  \BibitemOpen
  \bibfield  {author} {\bibinfo {author} {\bibfnamefont {S.~E.}\ \bibnamefont
  {{Gralla}}}\ and\ \bibinfo {author} {\bibfnamefont {A.}~\bibnamefont
  {{Lupsasca}}},\ }\href {\doibase 10.1103/PhysRevD.101.044031} {\bibfield
  {journal} {\bibinfo  {journal} {\prd}\ }\textbf {\bibinfo {volume} {101}},\
  \bibinfo {eid} {044031} (\bibinfo {year} {2020}{\natexlab{a}})},\ \Eprint
  {http://arxiv.org/abs/1910.12873} {arXiv:1910.12873 [gr-qc]} \BibitemShut
  {NoStop}%
\bibitem [{\citenamefont {{Lupsasca}}\ \emph {et~al.}(2024)\citenamefont
  {{Lupsasca}}, \citenamefont {{Mayerson}}, \citenamefont {{Ripperda}},\ and\
  \citenamefont {{Staelens}}}]{Lupsasca2024}%
  \BibitemOpen
  \bibfield  {author} {\bibinfo {author} {\bibfnamefont {A.}~\bibnamefont
  {{Lupsasca}}}, \bibinfo {author} {\bibfnamefont {D.~R.}\ \bibnamefont
  {{Mayerson}}}, \bibinfo {author} {\bibfnamefont {B.}~\bibnamefont
  {{Ripperda}}}, \ and\ \bibinfo {author} {\bibfnamefont {S.}~\bibnamefont
  {{Staelens}}},\ }\href {\doibase 10.48550/arXiv.2402.01290} {\bibfield
  {journal} {\bibinfo  {journal} {arXiv e-prints}\ ,\ \bibinfo {eid}
  {arXiv:2402.01290}} (\bibinfo {year} {2024})},\ \Eprint
  {http://arxiv.org/abs/2402.01290} {arXiv:2402.01290 [gr-qc]} \BibitemShut
  {NoStop}%
\bibitem [{\citenamefont {{Gurvits}}\ \emph {et~al.}(2022)\citenamefont
  {{Gurvits}} \emph {et~al.}}]{Gurvits2022}%
  \BibitemOpen
  \bibfield  {author} {\bibinfo {author} {\bibfnamefont {L.~I.}\ \bibnamefont
  {{Gurvits}}} \emph {et~al.},\ }\href {\doibase
  10.1016/j.actaastro.2022.04.020} {\bibfield  {journal} {\bibinfo  {journal}
  {Acta Astronautica}\ }\textbf {\bibinfo {volume} {196}},\ \bibinfo {pages}
  {314} (\bibinfo {year} {2022})},\ \Eprint {http://arxiv.org/abs/2204.09144}
  {arXiv:2204.09144 [astro-ph.IM]} \BibitemShut {NoStop}%
\bibitem [{\citenamefont {{Kurczynski}}\ \emph {et~al.}(2022)\citenamefont
  {{Kurczynski}} \emph {et~al.}}]{Kurczynski2022}%
  \BibitemOpen
  \bibfield  {author} {\bibinfo {author} {\bibfnamefont {P.}~\bibnamefont
  {{Kurczynski}}} \emph {et~al.},\ }in\ \href {\doibase 10.1117/12.2630313}
  {\emph {\bibinfo {booktitle} {Space Telescopes and Instrumentation 2022:
  Optical, Infrared, and Millimeter Wave}}},\ \bibinfo {series} {Society of
  Photo-Optical Instrumentation Engineers (SPIE) Conference Series}, Vol.\
  \bibinfo {volume} {12180},\ \bibinfo {editor} {edited by\ \bibinfo {editor}
  {\bibfnamefont {L.~E.}\ \bibnamefont {{Coyle}}}, \bibinfo {editor}
  {\bibfnamefont {S.}~\bibnamefont {{Matsuura}}}, \ and\ \bibinfo {editor}
  {\bibfnamefont {M.~D.}\ \bibnamefont {{Perrin}}}}\ (\bibinfo {year} {2022})\
  p.\ \bibinfo {pages} {121800M}\BibitemShut {NoStop}%
\bibitem [{\citenamefont {{Kudriashov}}\ \emph {et~al.}(2021)\citenamefont
  {{Kudriashov}} \emph {et~al.}}]{Kudriashov2021}%
  \BibitemOpen
  \bibfield  {author} {\bibinfo {author} {\bibfnamefont {V.}~\bibnamefont
  {{Kudriashov}}} \emph {et~al.},\ }\href {\doibase
  10.3724/SP.J.0254-6124.2021.0202} {\bibfield  {journal} {\bibinfo  {journal}
  {Chinese Journal of Space Science}\ }\textbf {\bibinfo {volume} {41}},\
  \bibinfo {pages} {211} (\bibinfo {year} {2021})},\ \Eprint
  {http://arxiv.org/abs/2105.06882} {arXiv:2105.06882 [astro-ph.IM]}
  \BibitemShut {NoStop}%
\bibitem [{\citenamefont {{Gralla}}(2020)}]{Gralla2020}%
  \BibitemOpen
  \bibfield  {author} {\bibinfo {author} {\bibfnamefont {S.~E.}\ \bibnamefont
  {{Gralla}}},\ }\href {\doibase 10.1103/PhysRevD.102.044017} {\bibfield
  {journal} {\bibinfo  {journal} {\prd}\ }\textbf {\bibinfo {volume} {102}},\
  \bibinfo {eid} {044017} (\bibinfo {year} {2020})},\ \Eprint
  {http://arxiv.org/abs/2005.03856} {arXiv:2005.03856 [astro-ph.HE]}
  \BibitemShut {NoStop}%
\bibitem [{\citenamefont {{Gralla}}\ and\ \citenamefont
  {{Lupsasca}}(2020{\natexlab{b}})}]{GrallaLupsasca2020c}%
  \BibitemOpen
  \bibfield  {author} {\bibinfo {author} {\bibfnamefont {S.~E.}\ \bibnamefont
  {{Gralla}}}\ and\ \bibinfo {author} {\bibfnamefont {A.}~\bibnamefont
  {{Lupsasca}}},\ }\href {\doibase 10.1103/PhysRevD.102.124003} {\bibfield
  {journal} {\bibinfo  {journal} {\prd}\ }\textbf {\bibinfo {volume} {102}},\
  \bibinfo {eid} {124003} (\bibinfo {year} {2020}{\natexlab{b}})},\ \Eprint
  {http://arxiv.org/abs/2007.10336} {arXiv:2007.10336 [gr-qc]} \BibitemShut
  {NoStop}%
\bibitem [{\citenamefont {{Paugnat}}\ \emph {et~al.}(2022)\citenamefont
  {{Paugnat}}, \citenamefont {{Lupsasca}}, \citenamefont {{Vincent}},\ and\
  \citenamefont {{Wielgus}}}]{Paugnat2022}%
  \BibitemOpen
  \bibfield  {author} {\bibinfo {author} {\bibfnamefont {H.}~\bibnamefont
  {{Paugnat}}}, \bibinfo {author} {\bibfnamefont {A.}~\bibnamefont
  {{Lupsasca}}}, \bibinfo {author} {\bibfnamefont {F.~H.}\ \bibnamefont
  {{Vincent}}}, \ and\ \bibinfo {author} {\bibfnamefont {M.}~\bibnamefont
  {{Wielgus}}},\ }\href {\doibase 10.1051/0004-6361/202244216} {\bibfield
  {journal} {\bibinfo  {journal} {\aap}\ }\textbf {\bibinfo {volume} {668}},\
  \bibinfo {eid} {A11} (\bibinfo {year} {2022})},\ \Eprint
  {http://arxiv.org/abs/2206.02781} {arXiv:2206.02781 [astro-ph.HE]}
  \BibitemShut {NoStop}%
\bibitem [{\citenamefont {{Vincent}}\ \emph {et~al.}(2022)\citenamefont
  {{Vincent}}, \citenamefont {{Gralla}}, \citenamefont {{Lupsasca}},\ and\
  \citenamefont {{Wielgus}}}]{Vincent2022}%
  \BibitemOpen
  \bibfield  {author} {\bibinfo {author} {\bibfnamefont {F.~H.}\ \bibnamefont
  {{Vincent}}}, \bibinfo {author} {\bibfnamefont {S.~E.}\ \bibnamefont
  {{Gralla}}}, \bibinfo {author} {\bibfnamefont {A.}~\bibnamefont
  {{Lupsasca}}}, \ and\ \bibinfo {author} {\bibfnamefont {M.}~\bibnamefont
  {{Wielgus}}},\ }\href {\doibase 10.1051/0004-6361/202244339} {\bibfield
  {journal} {\bibinfo  {journal} {\aap}\ }\textbf {\bibinfo {volume} {667}},\
  \bibinfo {eid} {A170} (\bibinfo {year} {2022})},\ \Eprint
  {http://arxiv.org/abs/2206.12066} {arXiv:2206.12066 [astro-ph.HE]}
  \BibitemShut {NoStop}%
\bibitem [{\citenamefont {{C{\'a}rdenas-Avenda{\~n}o}}\ and\ \citenamefont
  {{Lupsasca}}(2023)}]{CardenasAvendano2023}%
  \BibitemOpen
  \bibfield  {author} {\bibinfo {author} {\bibfnamefont {A.}~\bibnamefont
  {{C{\'a}rdenas-Avenda{\~n}o}}}\ and\ \bibinfo {author} {\bibfnamefont
  {A.}~\bibnamefont {{Lupsasca}}},\ }\href {\doibase
  10.1103/PhysRevD.108.064043} {\bibfield  {journal} {\bibinfo  {journal}
  {\prd}\ }\textbf {\bibinfo {volume} {108}},\ \bibinfo {eid} {064043}
  (\bibinfo {year} {2023})},\ \Eprint {http://arxiv.org/abs/2305.12956}
  {arXiv:2305.12956 [gr-qc]} \BibitemShut {NoStop}%
\bibitem [{\citenamefont {{Bardeen}}(1973)}]{Bardeen1973}%
  \BibitemOpen
  \bibfield  {author} {\bibinfo {author} {\bibfnamefont {J.~M.}\ \bibnamefont
  {{Bardeen}}},\ }in\ \href@noop {} {\emph {\bibinfo {booktitle} {Black Holes
  (Les Astres Occlus)}}},\ \bibinfo {editor} {edited by\ \bibinfo {editor}
  {\bibfnamefont {C.}~\bibnamefont {{Dewitt}}}\ and\ \bibinfo {editor}
  {\bibfnamefont {B.~S.}\ \bibnamefont {{Dewitt}}}}\ (\bibinfo  {publisher}
  {Gordon and Breach Science Publishers},\ \bibinfo {year} {1973})\ pp.\
  \bibinfo {pages} {215--239}\BibitemShut {NoStop}%
\bibitem [{\citenamefont {{Tiede}}\ \emph {et~al.}(2022)\citenamefont
  {{Tiede}}, \citenamefont {{Johnson}}, \citenamefont {{Pesce}}, \citenamefont
  {{Palumbo}}, \citenamefont {{Chang}},\ and\ \citenamefont
  {{Galison}}}]{Tiede2022}%
  \BibitemOpen
  \bibfield  {author} {\bibinfo {author} {\bibfnamefont {P.}~\bibnamefont
  {{Tiede}}}, \bibinfo {author} {\bibfnamefont {M.~D.}\ \bibnamefont
  {{Johnson}}}, \bibinfo {author} {\bibfnamefont {D.~W.}\ \bibnamefont
  {{Pesce}}}, \bibinfo {author} {\bibfnamefont {D.~C.~M.}\ \bibnamefont
  {{Palumbo}}}, \bibinfo {author} {\bibfnamefont {D.~O.}\ \bibnamefont
  {{Chang}}}, \ and\ \bibinfo {author} {\bibfnamefont {P.}~\bibnamefont
  {{Galison}}},\ }\href {\doibase 10.3390/galaxies10060111} {\bibfield
  {journal} {\bibinfo  {journal} {Galaxies}\ }\textbf {\bibinfo {volume}
  {10}},\ \bibinfo {pages} {111} (\bibinfo {year} {2022})},\ \Eprint
  {http://arxiv.org/abs/2210.13498} {arXiv:2210.13498 [astro-ph.HE]}
  \BibitemShut {NoStop}%
\bibitem [{\citenamefont {{Lockhart}}\ and\ \citenamefont
  {{Gralla}}(2022)}]{Lockhart2022}%
  \BibitemOpen
  \bibfield  {author} {\bibinfo {author} {\bibfnamefont {W.}~\bibnamefont
  {{Lockhart}}}\ and\ \bibinfo {author} {\bibfnamefont {S.~E.}\ \bibnamefont
  {{Gralla}}},\ }\href {\doibase 10.1093/mnras/stac2743} {\bibfield  {journal}
  {\bibinfo  {journal} {\mnras}\ }\textbf {\bibinfo {volume} {517}},\ \bibinfo
  {pages} {2462} (\bibinfo {year} {2022})},\ \Eprint
  {http://arxiv.org/abs/2208.09989} {arXiv:2208.09989 [astro-ph.HE]}
  \BibitemShut {NoStop}%
\bibitem [{\citenamefont {{Font}}(2008)}]{Font2008}%
  \BibitemOpen
  \bibfield  {author} {\bibinfo {author} {\bibfnamefont {J.~A.}\ \bibnamefont
  {{Font}}},\ }\href {\doibase 10.12942/lrr-2008-7} {\bibfield  {journal}
  {\bibinfo  {journal} {Living Reviews in Relativity}\ }\textbf {\bibinfo
  {volume} {11}},\ \bibinfo {eid} {7} (\bibinfo {year} {2008})}\BibitemShut
  {NoStop}%
\bibitem [{\citenamefont {{Porth}}\ \emph {et~al.}(2019)\citenamefont {{Porth}}
  \emph {et~al.}}]{Porth2019}%
  \BibitemOpen
  \bibfield  {author} {\bibinfo {author} {\bibfnamefont {O.}~\bibnamefont
  {{Porth}}} \emph {et~al.} (\bibinfo {collaboration} {Event Horizon Telescope
  Collaboration}),\ }\href {\doibase 10.3847/1538-4365/ab29fd} {\bibfield
  {journal} {\bibinfo  {journal} {\apjs}\ }\textbf {\bibinfo {volume} {243}},\
  \bibinfo {eid} {26} (\bibinfo {year} {2019})},\ \Eprint
  {http://arxiv.org/abs/1904.04923} {arXiv:1904.04923 [astro-ph.HE]}
  \BibitemShut {NoStop}%
\bibitem [{\citenamefont {{Wong}}\ \emph {et~al.}(2022)\citenamefont {{Wong}},
  \citenamefont {{Prather}} \emph {et~al.}}]{Wong2022}%
  \BibitemOpen
  \bibfield  {author} {\bibinfo {author} {\bibfnamefont {G.~N.}\ \bibnamefont
  {{Wong}}}, \bibinfo {author} {\bibfnamefont {B.~S.}\ \bibnamefont
  {{Prather}}},  \emph {et~al.},\ }\href {\doibase 10.3847/1538-4365/ac582e}
  {\bibfield  {journal} {\bibinfo  {journal} {\apjs}\ }\textbf {\bibinfo
  {volume} {259}},\ \bibinfo {eid} {64} (\bibinfo {year} {2022})},\ \Eprint
  {http://arxiv.org/abs/2202.11721} {arXiv:2202.11721 [astro-ph.HE]}
  \BibitemShut {NoStop}%
\bibitem [{\citenamefont {{Mizuno}}(2022)}]{Mizuno2022}%
  \BibitemOpen
  \bibfield  {author} {\bibinfo {author} {\bibfnamefont {Y.}~\bibnamefont
  {{Mizuno}}},\ }\href {\doibase 10.3390/universe8020085} {\bibfield  {journal}
  {\bibinfo  {journal} {Universe}\ }\textbf {\bibinfo {volume} {8}},\ \bibinfo
  {eid} {85} (\bibinfo {year} {2022})},\ \Eprint
  {http://arxiv.org/abs/2201.12608} {arXiv:2201.12608 [astro-ph.HE]}
  \BibitemShut {NoStop}%
\bibitem [{\citenamefont {{Akiyama}}\ \emph {et~al.}(2022)\citenamefont
  {{Akiyama}} \emph {et~al.}}]{EHT2022e}%
  \BibitemOpen
  \bibfield  {author} {\bibinfo {author} {\bibfnamefont {K.}~\bibnamefont
  {{Akiyama}}} \emph {et~al.} (\bibinfo {collaboration} {Event Horizon
  Telescope Collaboration}),\ }\href {\doibase 10.3847/2041-8213/ac6672}
  {\bibfield  {journal} {\bibinfo  {journal} {\apjl}\ }\textbf {\bibinfo
  {volume} {930}},\ \bibinfo {eid} {L16} (\bibinfo {year} {2022})}\BibitemShut
  {NoStop}%
\bibitem [{\citenamefont {{Wielgus}}\ \emph {et~al.}(2022)\citenamefont
  {{Wielgus}} \emph {et~al.}}]{Wielgus2022}%
  \BibitemOpen
  \bibfield  {author} {\bibinfo {author} {\bibfnamefont {M.}~\bibnamefont
  {{Wielgus}}} \emph {et~al.},\ }\href {\doibase 10.3847/2041-8213/ac6428}
  {\bibfield  {journal} {\bibinfo  {journal} {\apjl}\ }\textbf {\bibinfo
  {volume} {930}},\ \bibinfo {eid} {L19} (\bibinfo {year} {2022})},\ \Eprint
  {http://arxiv.org/abs/2207.06829} {arXiv:2207.06829 [astro-ph.HE]}
  \BibitemShut {NoStop}%
\bibitem [{\citenamefont {{Hudson}}\ \emph {et~al.}(2023)\citenamefont
  {{Hudson}}, \citenamefont {{Gurvits}}, \citenamefont {{Wielgus}},
  \citenamefont {{Paragi}}, \citenamefont {{Liu}},\ and\ \citenamefont
  {{Zheng}}}]{Hudson2023}%
  \BibitemOpen
  \bibfield  {author} {\bibinfo {author} {\bibfnamefont {B.}~\bibnamefont
  {{Hudson}}}, \bibinfo {author} {\bibfnamefont {L.~I.}\ \bibnamefont
  {{Gurvits}}}, \bibinfo {author} {\bibfnamefont {M.}~\bibnamefont
  {{Wielgus}}}, \bibinfo {author} {\bibfnamefont {Z.}~\bibnamefont {{Paragi}}},
  \bibinfo {author} {\bibfnamefont {L.}~\bibnamefont {{Liu}}}, \ and\ \bibinfo
  {author} {\bibfnamefont {W.}~\bibnamefont {{Zheng}}},\ }\href {\doibase
  10.1016/j.actaastro.2023.09.035} {\bibfield  {journal} {\bibinfo  {journal}
  {Acta Astronautica}\ }\textbf {\bibinfo {volume} {213}},\ \bibinfo {pages}
  {681} (\bibinfo {year} {2023})},\ \Eprint {http://arxiv.org/abs/2309.17127}
  {arXiv:2309.17127 [astro-ph.IM]} \BibitemShut {NoStop}%
\bibitem [{\citenamefont {{C{\'a}rdenas-Avenda{\~n}o}}\ \emph
  {et~al.}(2023)\citenamefont {{C{\'a}rdenas-Avenda{\~n}o}}, \citenamefont
  {{Lupsasca}},\ and\ \citenamefont {{Zhu}}}]{CardenasAvendano2022}%
  \BibitemOpen
  \bibfield  {author} {\bibinfo {author} {\bibfnamefont {A.}~\bibnamefont
  {{C{\'a}rdenas-Avenda{\~n}o}}}, \bibinfo {author} {\bibfnamefont
  {A.}~\bibnamefont {{Lupsasca}}}, \ and\ \bibinfo {author} {\bibfnamefont
  {H.}~\bibnamefont {{Zhu}}},\ }\href {\doibase 10.1103/PhysRevD.107.043030}
  {\bibfield  {journal} {\bibinfo  {journal} {\prd}\ }\textbf {\bibinfo
  {volume} {107}},\ \bibinfo {eid} {043030} (\bibinfo {year} {2023})},\ \Eprint
  {http://arxiv.org/abs/2211.07469} {arXiv:2211.07469 [gr-qc]} \BibitemShut
  {NoStop}%
\bibitem [{\citenamefont {{Lee}}\ and\ \citenamefont
  {{Gammie}}(2021)}]{Lee2021}%
  \BibitemOpen
  \bibfield  {author} {\bibinfo {author} {\bibfnamefont {D.}~\bibnamefont
  {{Lee}}}\ and\ \bibinfo {author} {\bibfnamefont {C.~F.}\ \bibnamefont
  {{Gammie}}},\ }\href {\doibase 10.3847/1538-4357/abc8f3} {\bibfield
  {journal} {\bibinfo  {journal} {\apj}\ }\textbf {\bibinfo {volume} {906}},\
  \bibinfo {eid} {39} (\bibinfo {year} {2021})},\ \Eprint
  {http://arxiv.org/abs/2011.07151} {arXiv:2011.07151 [astro-ph.IM]}
  \BibitemShut {NoStop}%
\bibitem [{\citenamefont {{Cunningham}}(1975)}]{Cunningham1975}%
  \BibitemOpen
  \bibfield  {author} {\bibinfo {author} {\bibfnamefont {C.~T.}\ \bibnamefont
  {{Cunningham}}},\ }\href {\doibase 10.1086/154033} {\bibfield  {journal}
  {\bibinfo  {journal} {\apj}\ }\textbf {\bibinfo {volume} {202}},\ \bibinfo
  {pages} {788} (\bibinfo {year} {1975})}\BibitemShut {NoStop}%
\bibitem [{\citenamefont {{Chael}}\ \emph {et~al.}(2021)\citenamefont
  {{Chael}}, \citenamefont {{Johnson}},\ and\ \citenamefont
  {{Lupsasca}}}]{Chael2021}%
  \BibitemOpen
  \bibfield  {author} {\bibinfo {author} {\bibfnamefont {A.}~\bibnamefont
  {{Chael}}}, \bibinfo {author} {\bibfnamefont {M.~D.}\ \bibnamefont
  {{Johnson}}}, \ and\ \bibinfo {author} {\bibfnamefont {A.}~\bibnamefont
  {{Lupsasca}}},\ }\href {\doibase 10.3847/1538-4357/ac09ee} {\bibfield
  {journal} {\bibinfo  {journal} {\apj}\ }\textbf {\bibinfo {volume} {918}},\
  \bibinfo {eid} {6} (\bibinfo {year} {2021})},\ \Eprint
  {http://arxiv.org/abs/2106.00683} {arXiv:2106.00683 [astro-ph.HE]}
  \BibitemShut {NoStop}%
\bibitem [{\citenamefont {{Guan}}\ \emph {et~al.}(2009)\citenamefont {{Guan}},
  \citenamefont {{Gammie}}, \citenamefont {{Simon}},\ and\ \citenamefont
  {{Johnson}}}]{Guan2009}%
  \BibitemOpen
  \bibfield  {author} {\bibinfo {author} {\bibfnamefont {X.}~\bibnamefont
  {{Guan}}}, \bibinfo {author} {\bibfnamefont {C.~F.}\ \bibnamefont
  {{Gammie}}}, \bibinfo {author} {\bibfnamefont {J.~B.}\ \bibnamefont
  {{Simon}}}, \ and\ \bibinfo {author} {\bibfnamefont {B.~M.}\ \bibnamefont
  {{Johnson}}},\ }\href {\doibase 10.1088/0004-637X/694/2/1010} {\bibfield
  {journal} {\bibinfo  {journal} {\apj}\ }\textbf {\bibinfo {volume} {694}},\
  \bibinfo {pages} {1010} (\bibinfo {year} {2009})},\ \Eprint
  {http://arxiv.org/abs/0901.0273} {arXiv:0901.0273 [astro-ph.GA]} \BibitemShut
  {NoStop}%
\bibitem [{\citenamefont {{Lu}}\ \emph {et~al.}(2023)\citenamefont {{Lu}} \emph
  {et~al.}}]{Lu2023}%
  \BibitemOpen
  \bibfield  {author} {\bibinfo {author} {\bibfnamefont {R.-S.}\ \bibnamefont
  {{Lu}}} \emph {et~al.},\ }\href {\doibase 10.1038/s41586-023-05843-w}
  {\bibfield  {journal} {\bibinfo  {journal} {\nat}\ }\textbf {\bibinfo
  {volume} {616}},\ \bibinfo {pages} {686} (\bibinfo {year} {2023})},\ \Eprint
  {http://arxiv.org/abs/2304.13252} {arXiv:2304.13252 [astro-ph.HE]}
  \BibitemShut {NoStop}%
\bibitem [{\citenamefont {{Shlentsova}}\ \emph {et~al.}(2024)\citenamefont
  {{Shlentsova}}, \citenamefont {{Roelofs}}, \citenamefont {{Issaoun}},
  \citenamefont {{Davelaar}},\ and\ \citenamefont {{Falcke}}}]{Shlentsova2024}%
  \BibitemOpen
  \bibfield  {author} {\bibinfo {author} {\bibfnamefont {A.}~\bibnamefont
  {{Shlentsova}}}, \bibinfo {author} {\bibfnamefont {F.}~\bibnamefont
  {{Roelofs}}}, \bibinfo {author} {\bibfnamefont {S.}~\bibnamefont
  {{Issaoun}}}, \bibinfo {author} {\bibfnamefont {J.}~\bibnamefont
  {{Davelaar}}}, \ and\ \bibinfo {author} {\bibfnamefont {H.}~\bibnamefont
  {{Falcke}}},\ }\href {\doibase 10.48550/arXiv.2403.03327} {\bibfield
  {journal} {\bibinfo  {journal} {arXiv e-prints}\ ,\ \bibinfo {eid}
  {arXiv:2403.03327}} (\bibinfo {year} {2024})},\ \Eprint
  {http://arxiv.org/abs/2403.03327} {arXiv:2403.03327 [astro-ph.HE]}
  \BibitemShut {NoStop}%
\end{thebibliography}%

\end{document}